\documentclass[reqno, 10pt]{amsart}

\usepackage{pdfsync}
\numberwithin{equation}{section}

\usepackage{enumerate}

\newcommand{\mc}[1]{{\mathcal #1}}
\newcommand{\mf}[1]{{\mathfrak #1}}

\newcommand{\bb}[1]{{\mathbb #1}}

\newcommand\BB{{\mathbb B}}

\newcommand\NN{{\mathbb N}}
\newcommand\RR{{\mathbb R}}

\newcommand\ZZ{{\mathbb Z}}

\usepackage{amssymb, amsmath}

\newtheorem{prop}{Proposition}
\newtheorem{theo}{Theorem}
\newtheorem{lemma}{Lemma}
\newtheorem{remark}{Remark}
\newtheorem{cor}{Corollary}

\title[]{}

\author{C\'edric Bernardin}
\address{%
Universit\'e de Lyon and CNRS, UMPA, UMR-CNRS 5669, ENS-Lyon,
46, all\'ee d'Italie, 69364 Lyon Cedex 07 - France.
}%
\author{Stefano Olla}
\address{%
CEREMADE, UMR CNRS 7534\\
Universit\'e Paris-Dauphine\\
75775 Paris-Cedex 16, France, \emph{and}\\
 INRIA - CERMICS, Projet MICMAC,\\
 Ecole des Ponts ParisTech\\ 
  6 \& 8 Av. Pascal, 77455 Marne-la-Vall\'ee Cedex 2, France\\
}%

\date{\today}

\thanks{\textsc{Acknowledgements.}  This paper has been partially
  suported by the European Advanced Grant {\em Macroscopic Laws and
    Dynamical Systems} (MALADY) (ERC AdG 246953), and by 
  French Ministry of Education through the ANR-10-BLAN 0108 grant. The authors thank P. Carmona and M. Hairer for useful discussions.}


\begin{document}


\title[Oscillators chain with random flip of velocities]{Transport
  Properties of a Chain of Anharmonic Oscillators with 
  random flip of velocities}

\maketitle

\begin{abstract}
We consider the stationary states of a chain of $n$ anharmonic coupled
oscillators, 
whose deterministic hamiltonian dynamics is perturbed by random
independent sign change of the velocities (a random mechanism that
conserve energy).  
The extremities are coupled to thermostats at different temperature 
$T_\ell$ and $T_r$  and subject to constant forces $\tau_\ell$ and 
$\tau_r$. 
If the forces differ $\tau_\ell \neq \tau_r$ the center of mass
of the system will move of a speed $V_s$ inducing a tension gradient
inside the system. 
Our aim is to see the influence of the tension gradient on the thermal
conductivity.  
We investigate the entropy production properties of the stationary
states, and we prove the existence of the Onsager matrix defined by
Green-kubo formulas (linear response). We also prove some explicit
bounds on the thermal conductivity, depending on the temperature.
\end{abstract}

 \section{Introduction}
 \label{sec:introduction}

Chains of anharmonic oscillators have been used as simple non-linear
microscopic models for the study of thermal conductivity. 
When coupled at the extremities to thermostats at different
temperatures, they have been the natural set-up, numerically and
theoretically, for the macroscopic Fourier law \cite{BRL}. 
When the interaction is anharmonic and a pinning potential is present,
the thermal conductivity is expected to be finite and generally
depending on the temperature. In fact a pinning potential destroys
translation invariance of the system (i.e. the conservation of
momentum), and temperature is the only parameter for equilibrium
states, corresponding to the energy conservation. If the chain is
unpinned, equilibrium states are parameterized also by the tension, and
we would expect a dependence of the thermal conductivity also on this
parameter.  
On the other hand in the unpinned case, we expect typically a
divergence of the thermal conductivity with the size of the system. 
 
We study here a stochastic perturbation of the dynamics of the
anharmonic unpinned oscillators, such that energy is conserved but not
momentum, but still the equilibrium measures are parameterized by
temperature and tension. This stochastic perturbation is extremely
simple: each particle waits independently an exponentially distributed time interval and then
flips the sign of its velocity. 

Furthermore, in order to produce a stationary state with a profile of
tension and of temperature, we apply at the extremities unequal
forces, and thermostats  at different temperatures.
In the corresponding stationary state the system will have a constant
energy current $J_s$ and a velocity $V_s$. These quantities are
related, at the first order, to the gradients of temperature and
tension by the Onsager matrix, that turns out to be diagonal. In fact
$V_s$ can be computed explicitely and is independent of the gradient
of temperature (in fact it is independent of the anharmonicity). 

While it is straightforward to show that $V_s$ is proportional to the
inverse of the size of the system, we are not able to prove the same
for $J_s$ (i.e. Fourier's law). Only in the harmonic case we are
able to show this property, by explicit calculations (cf Section
\ref{sec:harm}), together with the stationary flux of energy exchanged
with the thermostats. A closer look to these formulas reveals that, by
properly doing work on the system through the tension applied at the
boundaries, we can push energy into the hot reservoir, but it is
impossible to extract energy from the cold reservoir. In other words,
the system can act as a heater, but not as a refrigerator.

In Section \ref{sec:entprod} we study the entropy production of the
stationary state. In fact we relate what is known as \emph{entropy
  production} to the time derivative of the entropy of a nonstationary
state, in the following sense. Let $\mu_t$ a non stationary state of the
system at time $t$. Since we are dealing with a system with stochastic
thermostats, $\mu_t$ has a density with respect to the Lebesgue measure
and it can be defined its entropy $H(\mu_t)$. 
Computing the time derivative of $H(\mu_t)$, we find that it is
composed by two terms: the \emph{entropy rate of change} of the
thermostats and of the noise mechanism, that is always positive, and a
term that is usually called, up to a sign change, the entropy
production of the system in the state $\mu_t$. If we start with the
stationary state $\mu_{ss}$, then obviously the time derivative of the
entropy is zero, and the entropy production of the system equals the
entropy rate of change of the thermostats and the noise, i.e. is
positive. We prove that it is strictly positive unless we are in the
equilibrium state with same temperature and forces at the boundaries.
We actually work with a
local Gibbs measure as reference measure, but computations are similar.


\section{Mathematical formulation and main results}

\subsection{The model} 
 
We denote by $q_1, \dots, q_n$ the absolute positions of the
particles, and by $p_1, \dots, p_n$ the corresponding momenta
(particles mass is set equal to 1). The relevant coordinates are the
interparticle distances $r_x = q_x - q_{x-1}, x=2, \dots, n$. Thus,
the state space of 
our system is given by $\Omega_n={\mathbb R}^{n-1}\times {\mathbb R}^n$ and we shall denote a typical configuration $(r_2, \ldots,r_n, p_1, \ldots,p_n)$ by $\omega=(r,p) \in \Omega_n$. Between the
particles there is an anharmonic spring with potential $V(r_x)$, and
the corresponding hamiltonian dynamics is perturbed by independent
random flips of the sign of the velocities. 
Furthermore on the boundary particles $1$ and $n$ there are acting
Langevin thermostats at different temperature $T_\ell$ and $T_r$, and
two external constant forces $\tau_\ell$ and $\tau_r$.

The generator of the dynamics 
is given by  
$$
{\mc L}= {\mathcal A}_{\tau_\ell, \tau_r}  + \gamma {\mc S} +
\gamma_{\ell} {\mc B}_{1, T_\ell} + \gamma_r {\mc B}_{n, T_r}
$$
where ${\mc A}_{\tau_\ell, \tau_r}$ is the Liouville operator, 
${\mc B}_{j,T}$ the generator of the Langevin bath at temperature $T$
acting on the $j$--th particle and ${\mc S}$ the generator of the
noise. The strength of noise and thermostats are regulated by
$\gamma$, $\gamma_{\ell}$ and $\gamma_r$. The Liouville operator is
defined by 
\begin{equation}
  \label{eq:ss51}
   \begin{split}
    {\mathcal A }_{\tau_\ell, \tau_r}= 
    \sum_{x=2}^{n} \left(p_x - p_{x-1}\right) \partial_{r_x} +
        \sum_{x=2}^{n-1}\left(V'(r_{x+1}) - V'(r_{x})\right)
      \partial_{p_{x}}\\
    - \left(\tau_\ell - V'(r_2)\right) \partial_{p_1} 
    + \left(\tau_r - V'(r_n)\right) \partial_{p_n}.
  \end{split}
\end{equation}
The generators of the thermostats are given by
\begin{equation}
  \label{eq:ss49}
  {\mc B}_{j,T} = T \partial_{p_j}^2 - p_j \partial_{p_j}.
\end{equation}
The noise corresponds to independent velocity change of sign, i.e.
\begin{equation}
\label{eq:ss41}
({\mc S} f ) (\omega) = \sum_{x=2}^{n-1} \left(f(\omega^{x}) - f(\omega)\right), \quad f:\Omega_n \to \RR .
\end{equation}
Here, the configuration $\omega^{x}$ (resp. $p^x$) is the configuration obtained from $\omega$ (resp. $p$) by flipping the momentum of particle $x$, i.e. $\omega^x = (r,p^x)$ with $(p^x)_z =p_z, z \ne x, (p^x)_x=-p_x$.


In order to have a well defined process with good ergodic properties,
we assume that $V$ is a smooth even potential satisfying the following
assumptions:
\begin{enumerate}[(A)]
\item \label{eq:hyp} 
There exist positive constants $k \ge 2$ and $a_k >0$ such that
  \begin{equation*}
    \lim_{\lambda \to + \infty} \lambda^{-k} V (\lambda x) =a_k |x|^k,
    \quad \lim_{\lambda \to + \infty} \lambda^{1-k} V' (\lambda x) = k.
    a_k |x|^{k-1} {\rm sign} (x). 
  \end{equation*}
\item \label{eq:hypB}  
For any  $q \in \RR$, there exists $m=m(q) \ge 2$ such that $V^{(m)} (q) \ne 0$.
\end{enumerate}
Here ${\rm{sign}}(x)$ denotes the sign of $x$ and $V^{(m)}$ the
$m^{\rm{th}}$ derivative of $V$. 
Many of the results in the following should be valid for more
general potentials $V$, but this go beyond the porpouse of this
article.
\\

We shall denote by $(\omega (t))_{t \ge 0}$ the Markov process
generated by $\mathcal L$, and by  $(T_t)_{t\ge 0}$ the corresponding
semigroup, i.e. for any bounded function $f:\Omega_n \to \RR$, and any
$\omega \in \Omega_n$, 
\begin{equation*}
(T_t f) (\omega) = {\mathbb E}_{\omega} \left[ f (\omega (t)) \right] .
\end{equation*}
The energy of atom $x$ is defined by
\begin{equation*}
{\mc E}_1 =\cfrac{p_1^2}{2}, \quad {\mc E}_x =\cfrac{p_x^2}{2} + V(r_x), \quad x=2, \ldots,n.
\end{equation*}
For any positive constant $\theta$ we define the Lyapunov function
$W_\theta$ by 
\begin{equation*}
W_{\theta} (\omega) =\exp \left( \theta \sum_{x=1}^{n} {\mc E}_x \right), \quad  \omega \in \Omega_n,
\end{equation*} 
and the corresponding weighted Banach space $({\bb B}_{\theta}, \|
\cdot \|_{\theta})$: 
\begin{equation*}
\BB_{\theta} = \left\{ f: \Omega_n \to \RR \text{ continuous }, \; \|
  f\|_{\theta} = \sup_{\omega} \cfrac{| f(\omega)|}{W_{\theta}
    (\omega)} < +\infty \, \right\}.
\end{equation*}

In Section  \ref{sec:exist} is proved the following proposition:
\begin{prop}
\label{prop:car22}
Assume that $V$ satisfies (\ref{eq:hyp}). Then, if $\theta$ is
sufficiently small, the semigroup $(T_t)_{t \ge 0}$ can be extended to
a strongly Feller continuous semigroup on $\BB_{\theta}$ with a
probability transition that is absolutely continuous with respect to
the Lebesgue measure. Moreover, there exists a unique invariant
probability measure $\mu_{ss}$ for $(T_t)_{t\ge 0}$ and it is
absolutely continuous with respect to the Lebesgue measure.  
\end{prop}

\begin{remark}
  We believe that the density of $\mu_{ss}$ is smooth. The actual
  proof would require a delicate reworking of H\"ormander theorem.  
\end{remark}
 
When we are at equilibrium, i.e. $T_\ell=T_r=T$,
$\tau_\ell=\tau_r=\tau$, the generator ${\mc L}$ is denoted by ${\mc
  L}_{eq.}$. A simple computation shows that the Gibbs measure
$\mu^n_{\tau,T}$ with density w.r.t. the Lebesgue measure on
$\Omega_n$ given by 
\begin{equation*}
g^n_{\tau,T}  (r,p) = \prod_{x=1}^n \frac{e^{-\beta (\mathcal E_x - \tau
    r_x)}}{Z(\tau \beta, \beta)}, \quad \beta=T^{-1},
\end{equation*} 
 is invariant for ${\mc L}_{eq.}$. In the formula above we have
 introduced $r_1=0$ to avoid annoying notations. In fact, it is easy
 to check that ${\mc A}_{\tau,\tau}$ is antisymmetric in ${\mathbb
   L}^2 (\mu_{\tau,T}^n)$ and that ${\mc S}, {\mc B}_{j,T}$, $j=1,n$,
 are symmetric.

Let $\nabla$ be the discrete gradient defined, for any function $u:\ZZ
\to \RR$, by $(\nabla u) (x)= u(x+1)-u(x)$. The local conservation of
energy is expressed by the microscopic continuity equation 
\begin{equation*}
{\mc L} ({\mc E}_x) = -\nabla j_{x-1,x}, \quad x=1, \ldots, n,
\end{equation*}
where the energy current $j_{x,x+1}$ from site $x$ to site $x+1$ is
given by 
\begin{equation}
\begin{split}
j_{0,1} &= -\tau_{\ell} p_1 + \gamma_\ell (T_{\ell} -p_1^2), \\
j_{n,n+1} &=-\tau_r p_n -\gamma_r (T_r -p_n^2), \\
j_{x,x+1} &= - p_x V' (r_{x+1}), \, x=1,\ldots,n-1.
\end{split}
\end{equation}

The energy current $j_{0,1}$ (and similarly for $j_{n,n+1}$) is
composed of two terms: the term $-\tau_\ell p_1$ corresponds to the
work done on the first particle by the linear force and the term $
\gamma_\ell (T_{\ell} -p_1^2)$ is the heat current due to the left
reservoir.

We shall denote by $\langle \cdot \rangle_{ss}$ the expectation with
respect to the steady state. Let $V_s$ be the velocity of the center
of mass of the system and $J_{s}$ be the average energy current, which are defined by 
$$V_s=\langle p_x \rangle_{ss} \quad \text{and}\qquad 
J_s=\langle j_{x,x+1} \rangle_{ss} .
$$ 

Observe that we are in Lagrangian coordinates and that $J_s$ is really the interparticle exchange of energy, that does not take into account the trivial energy flux of the Eulerian coordinates due to the center of mass movement.

We have the simple relation between these two quantities
\begin{equation}
\label{eq:JV}
J_s = -\tau_\ell V_s + \gamma_\ell (T_\ell -\langle p_1^2 \rangle_{ss} ), \quad J_s = -\tau_r V_s - \gamma_r (T_r -\langle p_n^2 \rangle_{ss} ).
\end{equation}



The value of $V_s$ can be determined exactly and is independent of the nonlinearities present in the system. It follows that the tension profile, defined by $\tau_x = \langle V'(r_x) \rangle_{ss}$, is linear.
 
\begin{lemma}
\label{lem:velo}
The velocity $V_s$ of the center of mass is given by
\begin{equation}
\label{eq:V}
V_s = \cfrac{\tau_r -\tau_\ell}{2\gamma (n-2) + \gamma_\ell + \gamma_r}
\end{equation}
and the tension profile is linear:
\begin{equation}
  \label{eq:1}
  \tau_x = \cfrac{2\gamma (x-2) + \gamma_\ell}{2\gamma (n-2) + \gamma_\ell +
    \gamma_r} (\tau_r - \tau_\ell) + \tau_\ell,
\end{equation}
that implies
\begin{equation}
\lim_{n \to \infty} \tau_{[nu]} = 
\tau_\ell + (\tau_r -\tau_\ell) u, \quad u \in [0,1].
\end{equation}
\end{lemma}

\begin{proof}
We have 
\begin{equation}
\label{eq:pp}
\begin{cases}
{\mc L} (p_x) = (V' (r_{x+1}) -V' (r_x)) -2\gamma p_x, \quad x=2, \ldots,n-1,\\
{\mc L}  (p_1) = V' (r_2) -\tau_l -\gamma_\ell  p_1, \\
 {\mc L}  (p_n) = \tau_r -V' (r_n)  - \gamma_r p_n.
 \end{cases}
 \end{equation}
The first line of (\ref{eq:pp}) implies
\begin{equation*}
2 (n-2) \gamma V_s = \sum_{x=2}^{n-1} \left\{\tau_{x+1} -\tau_x \right\}= \tau_n -\tau_2.
\end{equation*}
The two last lines of (\ref{eq:pp}) give
\begin{equation*}
\tau_2 -\tau_\ell - \gamma_\ell V_s =0, \quad -\tau_n +\tau_r - \gamma_r V_s =0, 
\end{equation*}
and we get easily the value of $V_s$. 

To obtain the expression of the tension profile we observe that
(\ref{eq:pp}) gives a discrete difference equation which can be
solved exactly since we have the value of $V_s= \langle p_x
\rangle_{ss}$.
\end{proof}

For purely deterministic chain ($\gamma=0$), the
velocity $V_{s}$ is of order $1$, while the tension profile is flat
at the value $\left({\gamma_\ell + \gamma_r}\right)^{-1}\left[ {\gamma_\ell \tau_r + \gamma_r \tau_\ell }\right]$. The first effect of the
noise is to make $V_s$ of order $n^{-1}$ and to give a
normal tension profile.   

An immediate consequence of \eqref{eq:V} is that the mean velocity is
independent of the temperatures. Consequently, by Onsager symmetry, we
expect that the thermal conductivity at the first order is independent
of $\tau_r - \tau_l$, i.e. that the Onsager matrix is diagonal.
We actually prove this in Section \ref{sec:clr} as well as the
existence of the thermal conductivity by Green-Kubo formula (see
Section \ref{sec:GK}). 

In the next section we prove that the entropy production
$\Sigma(\mu_{ss})$ of the
stationary state is strictly positive if $T_\ell\neq T_r$ or $\tau_r
\neq \tau_l$, and is given by
\begin{equation*}
  \Sigma \left( \mu_{ss} \right) =\left(\cfrac{1}{T_r}-\cfrac{1}{T_\ell}
\right) J_s + \left(\cfrac{\tau_r}{T_r} -\cfrac{\tau_\ell}{T_\ell}
\right) V_s  .
\end{equation*}

In Section \ref{sec:bounds} we prove upper and lower bounds for the
thermal conductivity, in terms of the temperature. 

In Section~\ref{sec:harm} we show that for harmonic interactions, 
a Fourier's law holds and in particular we can compute the conductivity 
explicitely:
\begin{equation}
\lim_{n \to \infty} n J_s = \cfrac{1}{4\gamma} \left\{ (T_\ell -T_r)
  +(\tau_\ell^2 -\tau_r^2)\right\}. 
\end{equation}
Furthemore in this case can be computed explicitely the energy
currents between the thermostats and the system
(cf. \eqref{eq:jthermh}), and prove that 
the stationary state can increase the energy of the hottest
thermostat, but not lower the energy of the cold one.

The existence and uniqueness of the stationary state is proven in the
last Section \ref{sec:exist}. 

\section{Entropy production}
\label{sec:entprod}


Let $\beta_x$ (resp. $\tau_x$), $x=2, \ldots,n$, be the linear
interpolation profile between $T_{\ell}^{-1}=T^{-1}$ (resp. $\tau_\ell
=\tau$)  and $T_{r}^{-1} = (T +\delta T)^{-1}$ (resp. $\tau_r =\tau
-\delta \tau$). We  also define $\beta_1 =T^{-1}$.

We use now as reference probability measure the Gibbs local
equilibrium state $\mu_{lg} (dr,dp)= g(r,p) \,dr_2 \ldots dr_n\, dp_1
\ldots dp_n$ with 
\begin{equation}
  \label{eq:ss46tau}
  g (r,p) = \prod_{x=1}^n \frac{e^{-\beta_x (\mathcal E_x - \tau_x
    r_x)}}{Z(\tau_x\beta_x, \beta_x)}.
\end{equation}

Let $\mu_t=\mu_{lg} T_t $ be the law at time $t$ of the process
starting from $f_0 \cdot \mu_{lg}$ and ${\bar \mu}_{[s,s+t]} = t^{-1}
\int_s^{s+t} \mu_y dy $ be the time averaged law of the process on
$[s,s+t]$. The density of $\mu_t$ with respect to $\mu_{lg}$ is
denoted by $f_t$. This is a solution, in the sense of the
distributions, of the Fokker-Planck equation 
\begin{equation*}
\partial_t f_t = {\tilde {\mc L}}^* f_t
\end{equation*}
where ${\tilde{\mc L}}^{*} $ is the adjoint of ${\mc L}$ in ${\mathbb L}^2 (\mu_{lg})$.

The operators ${\mc B}_{1,T_\ell}, {\mc B}_{n,T_r}$ and ${\mc S}$ are
symmetric with respect  to $\mu_{lg}$ while the adjoint of the
Liouville operator is 
\begin{equation}
  \label{eq:ss49tau}
  \begin{split}
    {\mathcal A}_{\tau_\ell, \tau_r}^{*} &= -{\mathcal A }_{\tau_\ell, \tau_r}+ \sigma 
  \end{split}
\end{equation}
where
\begin{equation*}
\sigma= \sum_{x=1}^{n-1} \left(\beta_{x+1} - \beta_x\right) j_{x,x+1} + \sum_{x=2}^{n-1} \left(\beta_{x+1}\tau_{x+1} - \beta_x\tau_x\right) p_x .
\end{equation*}

Hence we get
\begin{equation}
\label{eq:adj}
\begin{split}
&{\tilde{\mc L}}^* = -{\mathcal A }_{\tau_\ell, \tau_r} +\gamma {\mc S} + \gamma_\ell  {\mc B}_{1,T_\ell} + \gamma_r  {\mc B}_{n,T_r}+\sigma ,\\
&{\mc L}+{\tilde {\mc L}}^* = 2\gamma {\mc S} + 2 \gamma_\ell  {\mc B}_{1,T_\ell} +2 \gamma_r {\mc B}_{n,T_r}+\sigma .
\end{split}
\end{equation}

We shall denote the relative entropy of a probability measure $\mu$
with respect to a probability measure $\nu$ by $H(\mu| \nu)$. This
is defined by  
\begin{equation}
H(\mu|\nu)= \sup_{\psi} \left\{ \int \psi d\mu -\log \left( \int e^{\psi} d\nu \right)\right\}
\end{equation}
where the supremum is carried over bounded measurable functions $\psi$. 
If $\frac{d\mu}{d\nu} = f$ exists and $\log f$ is $\mu$-integrable, we
have
\begin{equation*}
  H(\mu|\nu)= \int d\mu \log f .
\end{equation*}
So we can call  
$$
H(t)= \int f_t \log f_t d \mu_{lg}
$$ 
 the entropy of the system at time $t$. We choose $\mu_{lg}$ as a
 reference measure to estimate the entropy but a similar consideration
 could be performed by replacing $\mu_{lg}$ by the Lebesgue measure.  


\subsection{The smooth case}

We first give an informal argument to estimate the entropy production, which relies on smoothness properties of the density $f_{ss}$ (resp. $f_t$) of $\mu_{ss}$ (resp. $\mu_t$) with respect to $\mu_{lg}$. We have
\begin{equation}\label{dtent}
\begin{split}
\cfrac{d}{dt} H(t)&= \int \partial_t f_t \log f_t \, d \mu_{lg}
+ \partial_t \left( \int f_t \, d \mu_{lg} \right)\\ 
&= \int \left( {\tilde{\mc L}}^* f_t \right) \, \log f_t \, d\mu_{lg}=
\int f_t \, \left( {\tilde{\mc L}} \log f_t \right) \, d\mu_{lg}\\ 
&= \int {\mc A}_{\tau_\ell, \tau_r} f_t \, d\mu_{lg} \, +\,  \int f_t
\, \left[ \gamma {\mc S} + \gamma_{\ell} {\mc B}_{1,T_\ell} + \gamma_r
  {\mc B}_{n, T_r} \right] \left( \log f_t \right) \, d\mu_{lg}\\
&=\int \sigma f_t \, d\mu_{lg} \, + \, \int f_t \, \left[ \gamma {\mc
    S} + \gamma_{\ell} {\mc B}_{1,T_\ell} + \gamma_r{\mc B}_{n, T_r}
\right] \left( \log f_t \right) \, d\mu_{lg} 
\end{split}
\end{equation}
where we used (\ref{eq:ss49tau}) in the last equality. The first term
on the right-hand side is the entropy production of the hamiltonian
part of the dynamics, in the (non-stationary) state $f_t$,
\begin{equation*}
\Sigma (\mu_t) = -{\int} \sigma f_t d\mu_{lg} ,
\end{equation*}
 while the second term corresponds, up to the sign, to
the entropy production due to the thermostats and the
flipping noise. Notice that this second term is always positive.

If we start the system from the stationary state, $f_t=f_{ss}$ for all
$t$, and the left-hand side in \eqref{dtent} is equal to zero. Thus,
in the stationary state, we have 
\begin{equation*}
\Sigma (\mu_{ss}) = -\int f_{ss} \, \left[ \gamma {\mc S} +
  \gamma_{\ell} {\mc B}_{1,T_\ell} +\gamma_r {\mc B}_{n, T_r} \right]
\left( \log f_{ss} \right) \, d\mu_{lg}. 
\end{equation*}

By explicit calculation we have
\begin{equation*}
\begin{split}
\Sigma (\mu_{ss})
&= \gamma_\ell T_\ell \int \cfrac{(\partial_{p_1} f_{ss})^2 }{f_{ss}
} \, d\mu_{lg} +  \gamma_r T_r \int \cfrac{(\partial_{p_n} f_{ss})^2
}{f_{ss} } \,  d\mu_{lg} \\ 
&+ \cfrac{\gamma}{2}\,  \sum_{x=2}^{n-1}  \int \left[ f_{ss}
  (\omega^x) -f_{ss} (\omega)\right] \left[ \log f_{ss} (\omega^x)
  -\log f_{ss} (\omega) \right] \, d\mu_{lg} (\omega) ,
\end{split}
\end{equation*}
so that the entropy production of the stationary state is clearly
non-negative. In fact, it is strictly positive if the temperatures
$T_\ell$ and $T_r$ are different (see below).


%

\subsection{The non-smooth case}

Since we cannot prove that the density $f_{ss}$ is smooth, we have to
proceed in a different way. 

The entropy production $\Sigma (\alpha)$ of the probability measure $\alpha$ is given by 
\begin{equation}
\Sigma (\alpha) = \int \sigma (\omega)\, d\alpha (\omega).
\end{equation} 

The Dirichlet form ${\bb D} (\alpha)$ of a probability measure $\alpha$ with respect to the generator $\gamma {\mc S} +\gamma_\ell {\mc B}_{1, T_\ell} + \gamma_r {\mc B}_{n,T_r}$ is defined by
 \begin{equation*}
{\bb D} (\alpha) = \sup_{\psi} \left\{\, -\,  \int \cfrac{(2 \gamma {\mc S} + 2 \gamma_\ell {\mc B}_{1, T_\ell} + 2 \gamma_r {\mc B}_{n, T_r} ) {\psi} }{\psi}\,   d{\alpha} \right\}
\end{equation*} 
 where the supremum is carried over smooth functions $\psi$ bounded
 below by a positive constant and which are constant at infinity. It
 is easy to check that ${\bb D}$ is a positive convex and lower
 semicontinuous functional.

We first estimate the change of entropy.

 \begin{prop}
The relative entropy $H(\mu_t | \mu_{lg})$ is finite for every positive time $t$, and, for any $s,t \ge 0$, we have 
\begin{equation*}
\begin{split}
H(\mu_{t+s} | \mu_{lg}) -H(\mu_s |\mu_{lg}) &\le  - t \;  {\bb D} ( { \bar \mu}_{[s,s+t]})  + t \; \Sigma ({ \bar \mu}_{[s,s+t]} )
\end{split}
\end{equation*}
\end{prop}

\begin{proof}
Let us prove the claim for $s=0$, the general case being similar. In Section \ref{sec:exist} we prove that the semigroup $(T_t)_{t \ge 0}$ is such that the transition probabilities  have a positive density $q_t (\cdot, \cdot)$ with respect to the local equilibrium state $\mu_{lg}$. It follows that if $\nu=\psi \cdot  \mu_{lg}$ is a probability measure on $\Omega_n$ absolutely continuous with respect to $\mu_{lg}$ then $\nu T_t$ is also absolutely continuous w.r.t. $\mu_{lg}$ with a density that we denote by $\psi_t$. In fact, $\psi_t$ is given by
\begin{equation*}
\psi_t (\omega') = \int  q_t (\omega, \omega') \psi (\omega) g (\omega) d\omega.
\end{equation*}
Formally, $\psi_t$ is solution of the Fokker-Planck equation
\begin{equation*}
\partial_{t} \psi_t = {\tilde{\mc L}}^* \psi_t, \quad \psi_0 =\psi.
\end{equation*} 

If $\sigma$ was bounded, the solution $\psi_t$ to this equation would be given by the Feynman-Kac formula
\begin{equation}
\label{eq:FK}
{\psi_t} (\omega)={\mathbb E}_{\omega} \left[ \psi (\hat{\omega}(t)) \, e^{\int_0^t \sigma({\hat \omega} (s)) ds }\right]
\end{equation}
where $({\hat \omega} (t))_{t\ge 0}$ is the Feller process generated by $-{\mathcal A }_{\tau_\ell, \tau_r} +\gamma {\mc S} + \gamma_\ell  {\mc B}_{1,T_\ell} + \gamma_r  {\mc B}_{n,T_r}$. 

We claim that there exists $t_0>0$ such that (\ref{eq:FK}) makes sense if $t\le t_0$ and $\psi$ is bounded. It could be also possible to show the validity of (\ref{eq:FK}) for any time by assuming that $T_r -T_\ell, \tau_r -\tau_\ell$ are sufficiently small (see \cite{EPR}).

Observe there exists a constant $c_0:=c_0 (T_\ell, T_r, \tau_{\ell}, \tau_r)$ such that for any $\theta>0$,
$$e^{\theta | \sigma (\omega) |} \le W_{c_0 \theta} (\omega).$$
 By the bound (\ref{eq:ddd}) of Section \ref{sec:exist}, if $\theta$ is sufficiently small, then 
\begin{equation}
\sup_{t \ge 0} {\bb E}_{\omega} \left[ W_{c_0 \theta} ({\hat \omega} (t)) \right] \le C ( W_{c_0 \theta} (\omega) +1)
\end{equation}
for a positive constant $C(c_0, \theta)$. In fact, it is proved for $(\omega (t))_{t\ge 0}$ but the proof is the same for $({\hat \omega} (t))_{t\ge 0}$. If $t_0>0$ is sufficiently small, we have
\begin{equation*}
\begin{split}
{\mathbb E}_{\omega} \left[ e^{\int_0^t \sigma(\omega (s)) ds }\right]& \le \cfrac{1}{t} \int_0^t {\mathbb E}_{\omega} \left( e^{t \left| \sigma ({\hat \omega} (s)) \right| }\right) \,ds\\
&\le \sup_{s \le t_0} {\mathbb E}_{\omega} \left( e^{t_0 \left| \sigma ({\hat \omega} (s)) \right| }\right) \\
& \le \sup_{s \le t_0} {\mathbb E}_{\omega} \left[ W_{c_0 t_0} ({\hat \omega} (s)) \right]\\
&\le C'(t_0) \left[ W_{c_0 t_0} (\omega) +1\right] 
\end{split}
\end{equation*}
which proves the claim. Moreover, since $({\hat \omega} (t))_{t \ge
  0}$ defines a strongly continuous semigroup with generator
$-{\mathcal A }_{\tau_\ell, \tau_r} +\gamma {\mc S} + \gamma_\ell
{\mc B}_{1,T_\ell} + \gamma_r  {\mc B}_{n,T_r}$, we have 
\begin{equation}
\label{eq:genlstar}
\lim_{h \to 0} \cfrac{\psi_h -\psi_0}{h} = {\tilde{\mc L}}^* \psi_0 
\end{equation}
if $\psi_0$ is a smooth positive function constant at infinity.

Observe now that if $P,Q$ are two probability measures and $\phi$ is a density function w.r.t. $Q$  then
\begin{equation*}
H(P\, | \,Q)=H(P \,|\, \phi \cdot Q) +\int \log \phi \, dP .
\end{equation*}

We fix a time $0<h<t_0$ sufficiently small. For any probability measure $\alpha$ such that $H(\alpha| \mu_{lg}) < + \infty$, and any positive smooth function $\psi$ bounded bellow by a positive constant and constant at infinity, we have
\begin{eqnarray*}
H(\alpha T_{h} | \mu_{lg}) -H(\alpha| \mu_{lg})& =&H(\alpha T_{h}|  \psi_\tau \cdot \mu_{lg} ) + \int \log (\psi_{h})\, d( \alpha T_{h})-H(\alpha| \mu_{lg})\\ 
&=&H(\alpha T_{h}|  (\psi \cdot \mu_{lg}) T_h ) -H(\alpha| \mu_{lg}) + \int \log (\psi_{h}) d( \alpha T_{h})\\
&\le& H(\alpha |  (\psi \cdot \mu_{lg})) - H(\alpha|\mu_{lg}) + \int T_{h} \left( \log (\psi_{h}) \right) \, d\alpha\\
&\le& H(\alpha |  (\psi \cdot \mu_{lg})) - H(\alpha|\mu_{lg}) + \int \log \left( T_{h} \psi_{h}\right) \, d\alpha\\
&=& -\int \log \psi \, d\alpha +  \int \log \left( T_{h} \psi_{h} \right)\, d\alpha,
\end{eqnarray*}
where we used $H(\alpha T_h | \beta T_h ) \le H (\alpha | \beta)$, $\beta = \psi.\mu_{lg}$, in the first inequality and Jensen inequality in the second one. We write now $(T_{h} \psi_{h} )/ \psi = (T_{h} \psi_{h} -\psi )/\psi  + 1$ and we use the trivial inequality $\log (1+\eta) \le \eta$ to get
\begin{equation*}
H(\alpha T_{h} | \mu_{lg}) -H(\alpha | \mu_{lg}) \le \int \cfrac{T_{h} \psi_{h} -\psi }{\psi }\, d\alpha.
\end{equation*}

This shows in particular that $H(\alpha T_h | \mu_{lg}) <+ \infty$. By (\ref{eq:genlstar}), we have 
\begin{equation*}
T_{h} \psi_{h} -\psi = T_{h} (\psi_{h}-\psi) + (T_{h} \psi -\psi) = h ({\mc L} +{\tilde {\mc L}^*}) \psi + {h} {\varepsilon} ({h}, \psi)
\end{equation*}
where the remainder term ${\varepsilon} ({h}, \psi)$ vanishes as ${h}$ goes to $0$.

Fix a positive time $t$ and let $m$ be a positive integer sufficiently large. We define $h = t/m$ and we have
\begin{eqnarray*}
H(\alpha T_t | \mu_{lg}) -H(\alpha | \mu_{lg}) &=& \sum_{i=0}^{m-1} \left\{ H(\alpha T_{ (i+1)h} | \mu_{lg}) -H(\alpha T_{  i h} | \mu_{lg}) \right\}\\
& \le&  \cfrac{t}{m} \sum_{i=0}^{m-1} \int  \cfrac{ \left[({\mc L} +{\tilde {\mc L}^*}) \psi \right] }{\psi}\, d(\alpha T_{i h }) +t \varepsilon (h, \psi).
\end{eqnarray*}
As $m$ goes to infinity, the Riemann sum converges to 
\begin{equation*}
\int_{0}^t ds \, \int \cfrac{\left[({\mc L} +{\tilde {\mc L}^*}) \psi \right] }{\psi}\,  d(\alpha T_s)
\end{equation*} 
and the remainder term vanishes. Taking the infimum over functions $\psi$ and $\alpha=\mu_{lg}$, we get
\begin{equation*}
\begin{split}
H(\mu_t | \mu_{lg}) &\le t \inf_{\psi} \left\{ \int \cfrac{\left[({\mc L}+{\tilde {\mc L}^*}) \psi \right]}{\psi} d{\bar \mu}_{[0,t]}\right\}.
\end{split}
\end{equation*}
By using (\ref{eq:adj}), this concludes the proof.

\end{proof}

We recall that in the stationary state, $\langle j_{x,x+1} \rangle_{ss} =J_s$ and $\langle p_x \rangle_{ss} =V_s$ are independent of $x$.

\begin{theo}
\label{th:entprod}
The entropy production of the stationary state
\begin{equation*}
\Sigma \left( \mu_{ss} \right) = \left(\cfrac{1}{T_r}-\cfrac{1}{T_\ell}
\right) J_s + \left(\cfrac{\tau_r}{T_r} -\cfrac{\tau_\ell}{T_\ell}
\right) V_s  
\end{equation*} 
is strictly positive.  If $\tau_\ell =\tau_r$ it implies that the
energy flow goes from the hot reservoir to the cold reservoir. 
\end{theo}

\begin{proof}
Since the entropy is positive it follows that
\begin{equation}
\label{eq:impo}
0 \le {\mathbb D} ( {\bar \mu}_{[0,t]} ) \le \Sigma ({\bar \mu}_{[0,t]}).
\end{equation}

In the last section is proved the convergence as $t$ goes to infinity of
${\bar \mu}_{[0,t]}$ to the stationary state. Therefore, the entropy
production is non-negative 
\begin{equation*}
 \Sigma (\mu_{ss}) \ge 0. 
\end{equation*} 

If it is equal to zero then, by (\ref{eq:impo}), we have 
\begin{equation*}
\lim_{t \to \infty} {\bb D} ({\bar \mu}_{[0,t]}) =0
\end{equation*}
and by the lower semicontinuity of $\bb D$, we have
$${\bb D} (\mu_{ss})=0.$$ 
Recall that since $\mu_{ss}$ is the stationary state, we have
\begin{equation}
\label{eq:sss0}
\int ({\mc L} F)(\omega)\,  d\mu_{ss} (\omega) =0
\end{equation}
for any compactly supported smooth function $F$. We claim now that
\begin{equation}
\label{eq:DV}
\int \left( {\mc G} F \right) (\omega) \,  d\mu_{ss} (\omega) =0
\end{equation}
for $\mc G$ equal to ${\mc B} _{1,T_\ell}, {\mc B}_{n,T_r}, {\mc S}$. 

Indeed, since ${\bb D} (\mu_{ss})=0$, we have that for any smooth
function $\psi$ bounded by bellow and constant at infinity,  
\begin{equation}
\label{eq:inDV}
-\,  \int \cfrac{(2 \gamma {\mc S} + 2 \gamma_\ell {\mc B}_{1, T_\ell} + 2 \gamma_r {\mc B}_{n, T_r} ) {\psi} }{\psi}\,   d{\mu_{ss}} \; \le \; 0.
\end{equation}

Let us apply this with $\psi (\omega):=\psi_1 (p_1)$. Since ${\mc S}, {\mc B}_{n,T_r}$ does not act on $p_1$, it follows that the Dirichlet form with respect to ${\mc B}_{1,T_\ell}$ of the marginal $\mu_{ss}^1$ of $p_1$,
$$ \sup_{\psi_1} \left\{ -\,  \int \cfrac{ {\mc B}_{1, T_\ell} {\psi_1} }{\psi_1}\,   d{\mu_{ss}^1} \right\},$$
is negative, and hence equal to zero. It is well known that it implies $\mu_{ss}^1$ is the centered Gaussian law with variance $T_\ell$ (see \cite{DV}), and we get (\ref{eq:DV}) for ${\mc G }={\mc B}_{1,T_\ell}$. We similarly prove (\ref{eq:DV}) for ${\mc G} ={\mc B}_{n,T_r}$. By applying (\ref{eq:inDV}) with a function $\psi (\omega):={\bar \psi} (p_2, \ldots,p_{n-1})$, we get that the Dirichlet form of the marginal ${\bar{\mu}}_{ss}$ of $(p_2, \ldots,p_{n-1})$ with respect to ${\mc S}$,
 $$\bar {{\bb D}} ({\bar \mu}_{ss})= \sup_{{\bar \psi}} \left\{ -\,  \int \cfrac{ {\mc S} {\bar \psi}}{{\bar \psi} }\,   d {\bar{\mu}}_{ss} \right\},$$
 is negative, and hence equal to zero. Let ${\bar{f_{ss}}}$ be the density of $\bar{\mu_{ss}}$ with respect to the Lebesgue measure. It is easy to show that 
 \begin{equation*}
 {\bar {\bb D}} ({ \bar{{\mu}}_{ss} })=  \cfrac{1}{2} \int_{\RR^{n-2}} \left[ \sqrt{ {\bar f}_{ss} (p^x) }-\sqrt{{\bar f}_{ss} (p)}\right]^2 dp_2 \ldots dp_{n-1}.
 \end{equation*}
It follows that ${\bar \mu}_{ss}$ is invariant by any  flip $p \to p^x$, $x=2, \ldots,n-1$, and this concludes the proof of (\ref{eq:DV}). 

By (\ref{eq:sss0}) and (\ref{eq:DV}), we get that for any smooth compactly supported function $F$,
\begin{equation*}
\int \left( {\mc A}_{\tau_\ell, \tau_r}  F \right) (\omega)\,  d\mu_{ss} (\omega) =0, 
\end{equation*}  
and consequently that
\begin{equation*}
\int \left( {\mc L}_0 F  \right) (\omega) \, d\mu_{ss} (\omega) =0,
\end{equation*}   
where 
\begin{equation}
\label{eq:L0}
{\mc L}_0=  {\mc A}_{\tau_\ell, \tau_r} + \gamma_\ell  {\mc B} _{1,T_\ell} + \gamma_r {\mc B}_{n,T_r}
\end{equation}
is the generator of the deterministic chains in contact with the  two heat baths. Thus, $\mu_{ss}$ is equal to the (unique) invariant probability measure $\nu_{ss}$ of the process generated by ${\mc L}_{0}$. By using a similar argument to \cite{EPR}, one can show that 
\begin{equation*}
\Sigma (\nu_{ss}) =0 
\end{equation*}
implies $T_\ell = T_r$. Hence, if $T_\ell \ne T_r$, the entropy production of the stationary state $\mu_{ss}$ is strictly positive.

\end{proof}

From the equality $J_s = \langle j_{0,1} \rangle_{ss} =\langle j_{n, n+1} \rangle_{ss} $ we get
\begin{equation*}
\gamma_\ell \langle p_1^2 \rangle_{ss} + \gamma_r \langle p_n^2 \rangle_{ss} = \gamma_\ell T_\ell + \gamma_r T_r + \cfrac{(\tau_r -\tau_\ell)^2}{2 \gamma (n-2)  + \gamma_\ell +\gamma_r}.
\end{equation*}
This shows there exists a constant $C$ depending on the parameters of the model ($T_\ell, T_r, \tau_\ell, \tau_r , \gamma_\ell, \gamma_r$) but not on $n$ such that
\begin{equation}
\label{eq:ubp}
\langle p_1^2 \rangle_{ss} \le C, \quad \langle p_n^2 \rangle_{ss} \le C.
\end{equation}

It is expected there exists a positive constant $C$ independent of the size $n$ such that $\langle {\mc E}_x \rangle_{ss} \le C$ for any $x=1, \ldots,n$. Unfortunately, apart from the harmonic case discussed in Section \ref{sec:harm}, we do not know how to prove such a bound. 

%




\section{Conductivity: Linear response}
\label{sec:clr}

In this section, the discussion is kept at some informal level of
mathematical rigor.  We shall denote by ${\tilde f_{ss}}$ the
derivative of the stationary state $\mu_{ss}$ with respect to the
local equilibrium state $\mu_{lg}$.  It is solution, in the sense of the distributions, of the equation
\begin{equation}
 \label{eq:ss48}
{\tilde{\mc L}}^{*} \, \tilde f_{ss} = 0.
\end{equation}

We assume that $T_r=T +\delta T, T_\ell= T$ and $\tau_r = \tau -\delta \tau, \tau_\ell= \tau$ with $\delta T, \delta \tau$ small. Recall (\ref{eq:adj}). At first order in $\delta T$ and $\delta \tau$, we have
\begin{equation*}
  \begin{split}
    \tilde {\mc L}^{*} & = -{\mathcal A}_{\tau_\ell, \tau_r} + \gamma {\mc S} + {\mc B}_{1, T} + {\mc B}_{n, T+ \delta T} \\
    &- \frac {\delta T}{T^2 n} \sum_{x=1}^{n-1} \left(j_{x,x+1} + \tau p_x \right) -\frac{\delta \tau}{nT} \sum_{x=1}^{n-1} p_x + o(\delta T, \delta \tau)\\
    &=  -{\mathcal A}_{\tau_\ell,\tau_r} +\gamma {\mc S}+ {\mc B}_{1, T_\ell} + {\mc B}_{n, T_r}\\
    &-\frac {\delta T}{T^2 n} \sum_{x=1}^{n-1} \left(j_{x,x+1} + \tau p_x \right) -\frac{\delta \tau}{nT} \sum_{x=1}^{n-1} p_x  + \gamma_r \delta T \partial_{p_n}^2 + o(\delta T, \delta \tau)\\
    &=  {\mc L}_{eq.}^{*} +  \gamma_r \delta T \partial_{p_n}^2 -\delta \tau \partial_{p_n} -\frac {\delta T}{T^2 n} \sum_{x=1}^{n-1} \left(j_{x,x+1} + \tau p_x \right) +\frac{\delta \tau}{nT} \sum_{x=1}^{n-1} p_x+ o(\delta T, \delta \tau)
  \end{split}
\end{equation*}
where $ {\mc L}_{ eq.}^{*}=-{ \mathcal A}_{\tau,\tau} + \gamma {\mc S} + \gamma_\ell {\mc B}_{1, T} + \gamma_r {\mc B}_{n, T} $ is the adjoint in ${\mathbb L}^{2} (\mu^n_{\tau,T})$ of
\begin{equation}
  \label{eq:ss43}
   {\mc L}_{eq.} ={ \mathcal A}_{\tau,\tau} + \gamma {\mc S} + \gamma_\ell {\mc B}_{1, T} + \gamma_r {\mc B}_{n, T} .
\end{equation}

We now expand $\tilde f_{ss}$ at the linear order in $\delta T$ and $\delta \tau$:
\begin{equation}
  \label{eq:ss55tau}
  \tilde f_{ss} = 1 + \tilde u \, \delta T +\tilde v \,  \delta \tau +
  o(\delta T, \delta \tau)
\end{equation}
and we get that $\tilde u$ and $\tilde v$ are solution of
\begin{equation}
  \label{eq:ss51}
  \begin{split}
    \mc L_{eq.}^{*} \tilde u &= \frac 1{T^2 n} \sum_{x=1}^{n-1}
    \left(j_{x,x+1} + \tau p_x \right),
    \\
     \mc L_{eq.}^{*} \tilde v &= \frac{1}{nT}
     \sum_{x=1}^{n-1} p_x . 
  \end{split}
\end{equation}

We can now compute the average energy current at the first order in
$\delta T$ and $\delta \tau$ in the thermodynamic limit $n \to \infty$. 

Before, we need to introduce some  notations. Let $\Omega= \cap_{\alpha >0} \Omega_\alpha$ where $\Omega_{\alpha}$ is the Banach space composed of configurations $\omega=(r,p) \in (\RR \times \RR)^{\ZZ}$ such that the norm $\| \omega \|_{\alpha}$ defined by $\| \omega \|_{\alpha}^2 =\sum_{x \in \ZZ} e^{-\alpha |x|} \left[ p_{x}^2 + r_{x}^2 \right]$ is finite. We equip $\Omega$ with the topology induced by these norms. If the potential $V$ is such that $V'' \le C$ then one can prove that the infinite dynamics is well defined for any initial condition belonging to $\Omega$ and in particular on a set of initial conditions of full probability with respect to any  infinite volume Gibbs measure $\mu_{\tau,T}$ with temperature $T$ and pressure $\tau$ (\cite{BO}, \cite{FFL}). As in the finite dimensional case, for any configuration $\omega=(r,p) \in \Omega$, the configuration $\omega^x = (r,p^x)$ is obtained from $\omega$ by flipping the momentum of particle $x$. We shall denote by $C_0^{k} (\Omega)$ the space composed of compactly supported local functions on $\Omega$ which are differentiable up to order $k$, $k \ge 1$. The generator of the infinite dynamics is given by ${\mf L} ={\mf A} +\gamma {\mf S}$ where, for any $f \in C_0^1 (\Omega)$,
\begin{equation*}
\begin{split}
&({\mf A} f)(\omega) =  \sum_{x \in \ZZ} \left[ \left(p_x - p_{x-1}\right) \partial_{r_x} f + \left(V'(r_{x+1}) - V'(r_{x})\right) \partial_{p_{x}} f \right] (\omega),\\
&({\mf S} f) (\omega) = \sum_{x \in \ZZ} \left[ f(\omega^x) -f (\omega) \right].
\end{split}
\end{equation*}

 We denote by $\theta_x: \Omega \to \Omega$ the shift by $x$: for any $\eta \in \Omega$, $(\theta_x \eta)_{z} =\eta_{x+z}$; for any $g: \Omega \to \RR$, $(\theta_x g)(\eta) = g(\eta_x \eta)$. Let ${\mc H}_{\tau,T}$ be the completion w.r.t. the semi-inner product $\ll \cdot, \cdot \gg$ defined for local functions $f,g : \Omega \to \RR$, by 
\begin{equation}
\ll f, g \gg = \sum_{x \in \ZZ} \left\{ \mu_{\tau,T} (f \theta_x g) - \mu_{\tau,T} (f) \mu_{\tau,T} (g)\right\}.
\end{equation}
Observe that in ${\mc H}_{\tau,T}$ every discrete gradient $\theta_1 f - f$ is equal to zero.

Let ${\hat J}_s$ and ${\hat V_s}$ be the limiting average energy current and velocity:
\begin{equation}
{\tilde J}_s = \lim_{n \to \infty} n \langle j_{0,1} \rangle_{ss}, \quad {\hat V}_{s} = \lim_{n \to \infty} n \langle p_0 \rangle_{ss},
\end{equation}
and define ${\hat J}_{s} = {\tilde J}_s + \tau {\hat V}_s$. We expect that as $n$ goes to infinity and, at first order in $\delta T$ and $\delta \tau$,
\begin{equation*}
\left(
\begin{array}{c}
{\hat J}_s\\
{\hat V}_s
\end{array}
\right)
= - \, 
\kappa (T,\tau) \, 
\left(
\begin{array}{c}
\delta T\\
\delta \tau
\end{array}
\right)
\end{equation*}
with 
\begin{equation*}
\kappa (T,\tau)=
\left(
\begin{array}{cc}
\kappa^e& \kappa^{e,r}\\
\kappa^{r,e}&\kappa^r
\end{array}
\right)
\end{equation*}
the {\textit{ thermal conductivity}} matrix. By (\ref{eq:ss51}), we get that in the thermodynamic limit $n \to \infty$,
\begin{equation}
\label{eq:ke}
\begin{split}
\kappa^e = T^{-2} \ll j_{0,1} +\tau p_0\,,\, (-{\mf L})^{-1}\, (j_{0,1} +\tau p_0) \gg,\\
\kappa^{e,r} =-T^{-1} \ll p_0\,,\, (-{\mf L})^{-1} \,( j_{0,1}+\tau p_0) \gg, 
\end{split}
\end{equation} 
and
\begin{equation}
\label{eq:kr}
\begin{split}
\kappa^r =T^{-1} \ll p_{0} \,,\,(- {\mf L})^{-1}\, (p_0) \gg,\\
\kappa^{r,e} =-T^{-2} \ll j_{0,1} + \tau p_0\,,\, (-{\mf L})^{-1} \,( p_0) \gg. 
\end{split}
\end{equation}

The argument above is formal. In fact even proving the existence of the transport coefficients defined by (\ref{eq:ke}), (\ref{eq:kr}) is a non-trivial task. It can be made rigorous for ${\hat V}_s$ since we have the exact expression of $V_s$. From Lemma \ref{lem:velo}, we have, even if $\delta \tau ,\delta T$ are not small,
$${\hat V}_s = -\cfrac{\delta \tau}{2\gamma}$$
If the formal expansion can be made rigorous, the quantities $\kappa^r, \kappa^{r,e}$, defined by (\ref{eq:kr}), shall satisfy
\begin{equation}
\label{eq:krr}
{\kappa^r} = (2 \gamma)^{-1}, \quad {\kappa}^{r,e} =0.
\end{equation}  

In Theorem \ref{th:GK} we show that the transport coefficients defined in (\ref{eq:ke}), ({\ref{eq:kr}) exist. In Proposition \ref{prop:ons} we prove (\ref{eq:krr}) and the Onsager relations 
\begin{equation}
\kappa^{r,e} =\kappa^{e,r} \; (=0).
\end{equation}

Consequently, if $\delta T$ and $\delta \tau$ are small and of the
same order, the system can not be used as a refrigerator or a
boiler: at the first order, a gradient of tension does
not contribute to the energy current. 
The argument above says nothing about the possibility to
realize a heater or a refrigerator if $\delta \tau $ is not of the
same order as $\delta T$. For the harmonic chain, we will see in
Section \ref{sec:harm} that it is possible to get a heater if $\delta
\tau $ is of order $\sqrt {\delta T}$.

\section{Existence of the Green-Kubo formula}
\label{sec:GK}

To simplify notations we denote ${\mathcal
  H}_{\tau,T}$ by ${\mathcal H}$ (defined in Section \ref{sec:clr}).  We assume that the unbounded
operator ${\mf L}$ with domain ${\mc D} ({\mf L})$ is the generator of
a strongly continuous positive semigroup on ${\mc H}$ and that the set $C_0^1 (\Omega)$ forms a dense subset of ${\mc D} (\mf L)$. Similarly, we assume that ${\mf S}$ (resp. $\mf A$) with domain
denoted by ${\mc D} ({\mf S})$ (resp. ${\mc D} (\mf A)$) is the
generator of a strongly continuous positive semigroup on ${\mc H}$ and
that $C_0^1 (\Omega)$ forms a dense subset
of ${\mc D} (\mf S)$ (resp. ${\mc D}({\mf A})$). We have ${\mc D} (\mc
L) \subset {\mc D} (\mf S) \cap {\mc D} (\mf A)$. The generator ${\mf L}$ has the decomposition ${\mf A} +\gamma { \mf S}$ in its antisymmetric and symmetric part in $\mc H$. These assumptions can be proved without difficulty in the case $V''$ uniformly bounded (\cite{BO}, \cite{FFL}). They should be true for more general potentials but proofs should be quite technical (see \cite{F1}).

Let $\chi$ be the set of functions $\xi:\ZZ \to \NN$ a.s. equal to zero. For a given $\xi \in \chi$ we denote by $H_{\xi} (p)$ the polynomial function
\begin{equation*}
H_{\xi} (p) = \prod_{x \in \ZZ} h_{\xi_x} (p_x)
\end{equation*}
where $(h_n)_{n \ge 0}$ are the normalized Hermite polynomials w.r.t. the centered one dimensional Gaussian measure with variance $T$. It is well known that $(H_\xi)_{\xi \in \chi}$ forms an orthonormal basis of the Hilbert space composed of square integrable functions with respect to the product centered  Gaussian measures with variance $T$. It follows that every functions $f \in {\mc H}$ can be decomposed in the form
\begin{equation*}
f(r,p) = \sum_{\xi \in \chi} F(\xi,r) H_{\xi} (p).
\end{equation*}

Let ${\mc H}^a$ (resp. ${\mc H}^s$) be the set of functions $f:\Omega \to \RR$ antisymmetric (resp. symmetric) in $p$, i.e. $f(r,p)=-f(r,-p)$ (resp. $f(r,p)=f(r,-p)$)  for every configuration $(r,p) \in \Omega$. For example, the functions $j_{0,1}$, $p_0$ and every linear combination of them are antisymmetric in $p$. 

Since the Hermite polynomial $h_n$ is even if $n$ is even and odd
otherwise, the space ${\mc H}^a$ (resp. ${\mc H}^s$)
coincides with the set of functions $f = \sum_{\xi \in \chi_1} F(\xi, r)
H_{\xi}$ (resp. $f = \sum_{\xi \in \chi_0 } F(\xi, r) H_{\xi} $) where 
\begin{equation*}
\chi_1= \left\{ \xi \in \chi\, ;\, |\xi| {\text{ is odd}} \right\}, \quad \chi_0= \left\{ \xi \in \chi\, ;\, |\xi| {\text{ is even}} \right\}
\end{equation*}
with $|\xi| = \sum_x \xi_x$.

The system is conservative and does not have a spectral gap but we have a similar property for the antisymmetric functions.

 \begin{lemma}
 \label{lem:sg}
The noise operator ${\mf S}$ lets ${\mc H}^a$ and ${\mc H}^s$ invariant. For any function $f \in {\mc D} (\mf S) \cap {\mc H}^a$ we have
\begin{equation}
\label{eq:sg1}
 2 \ll f, f \gg \; \le \; \ll f, -{\mf S} f \gg.    
\end{equation}
Moreover, for any local function $f \in {\mc H}^a$, there exists a local function $h \in {\mc H}^a$ such that
$${\mf S}h =f.$$ 
\end{lemma} 

\begin{proof}

Let $f$ be a local function belonging to ${\mc H}$ with decomposition given by
$$f = \sum_{\xi \in {\chi}} F(\xi, r) H_{\xi}(p). $$
For $\xi \in \chi$ we note $W(\xi)$ the number defined by $W(\xi)=|\{ x \in \ZZ\; ; \; \xi_x \ge 1 \text{ is odd }\}|$. Observe that $W(\xi) \ge 1$ as soon as $\xi \in \chi_1$.
We have
\begin{equation}
\label{eq:dec}
{\mf S} (f) = -2 \sum_{\xi \in \chi}  W(\xi) F(\xi, r) H_\xi (p) 
\end{equation}
because ${\mf S} (h_{n} (p_x))= ((-1)^n -1) h_{n} (p_x)$. This shows that ${\mc H}^a$ and ${\mc H}^s$ are invariant by ${\mf S}$.

Observe that for any centered local functions $f$ and $g$,
\begin{equation*}
\ll f, g \gg =\lim_{k \to \infty} \mu_{\tau,T} \left( \cfrac{1}{\sqrt{2k+1}} \sum_{|x| \le k} \theta_x f \, , \, \cfrac{1}{\sqrt{2k+1}} \sum_{|x| \le k} \theta_x g\right)
\end{equation*}
and that ${\mf S} (\theta_x f) = \theta_x ({\mf S} f)$. Thus, the proof of (\ref{eq:sg1}) is reduced to show that if $f \in {\mc H}^a$ then 
\begin{equation*}
2 \mu_{\tau, T} (f^2) \le \mu_{\tau,T} (( -{\mf S} f) f).
\end{equation*}

Since $f \in {\mc H}^a$, we have $F(\xi,r)=0$ if $\xi \notin \chi_1$.  By (\ref{eq:dec}) we have to prove
\begin{equation*}
2 \sum_{\xi \in \chi_1} \mu_{\tau , T} (F^2(\xi,r))  \le 2 \sum_{\xi \in \chi_1} W(\xi) \mu_{\tau , T} (F^2(\xi,r))
\end{equation*} 
which is valid since $W(\xi) \ge 1$ as soon as $\xi \in \chi_1$.

The second claim of the proposition follows from (\ref{eq:dec}) by taking $h$ given by
\begin{equation*}
h(r,p)=\sum_{\xi \in \chi_1} H(\xi, r)H_{\xi} (p), \quad H(\xi, r)= (-2 W(\xi))^{-1} F(\xi,r).
\end{equation*}

\end{proof}

\begin{theo}
\label{th:GK}
Let $f,g \in {\mc H}_a$. Then, the limit 
\begin{equation*}
\sigma(f, g) =\lim_{\lambda \to 0} \ll f\, ,\, (\lambda - {\mf L})^{-1} \, g \gg
\end{equation*}
exists and $\sigma(f,g)=\sigma(g,f)$.
\end{theo}

\begin{proof}
We introduce the $H_1$ norm corresponding to the symmetric part ${\mf S}$ of ${\mf L}$
\begin{equation*}
\| u \|_1^2 = \ll u, (-{\mf S}) u \gg
\end{equation*} 
and ${\mathcal H}_1$ the Hilbert space obtained by the completion of ${\mathcal H}$ w.r.t. this norm. The corresponding scalar product is denoted by $\ll \cdot, \cdot \gg_1$.

By density of local functions in ${\mc H}$ we can assume that $f,g$ are local functions.  Let $u_\lambda$ be the solution of the resolvent equation 
\begin{equation}
\label{eq:res}
\lambda u_\lambda -{\mf L} u_\lambda = g.
\end{equation} 
 We multiply (\ref{eq:res}) by $u_\lambda$ and integrate w.r.t. $\ll\cdot,\cdot\gg$ and we get
 \begin{equation*}
 \lambda \ll u_\lambda, u_\lambda \gg +\gamma \| u_\lambda \|_1^2 = \ll u_\lambda, g \gg.
 \end{equation*}
 Since $g \in {\mc H}^a$ we have by lemma \ref{lem:sg} there exists a local function $h\in {\mc H}^a$ such that ${\mf S} h = g$. By Schwarz inequality, we have
 \begin{equation*}
 \| u_\lambda \|_1^2 \leq C^2 \gamma^{-1}
 \end{equation*}
 and
 \begin{equation*}
 \lambda \ll u_\lambda, u_\lambda \gg \leq C^2 \gamma^{-1}.
 \end{equation*}
Since $(u_\lambda)_{\lambda}$ is a bounded sequence in ${\mathcal H}_1$, we can extract a weakly converging subsequence in ${\mathcal H}_1$. We continue to denote this subsequence by $(u_\lambda)_\lambda$ and we denote by $u_0$ the limit.

Let $u_\lambda (p,r)= u_\lambda^s (p,r) + u_{\lambda}^a (p,r)$ be the decomposition of $u_\lambda$ in its symmetric and antisymmetric part in the $p$'s. Since $g$ is antisymmetric in the $p$'s, we have that $\ll u_\lambda, g \gg = \ll u_\lambda^a , g \gg$. Furthermore ${\mf S}$ preserves the parity in $p$  (see lemma \ref{lem:sg}) while it is inverted by ${\mf A}$. We have the following decomposition
\begin{eqnarray*}
\lambda u_{\lambda}^s - \gamma {\mf S} u_{\lambda}^s -{\mf A} u_{\lambda}^a =0,\\
\mu u_{\mu}^a - \gamma {\mf S} u_{\mu}^a -{\mf A} u_{\mu}^s =g.
\end{eqnarray*} 

We multiply the first equality  by $u_\mu^s$ and the second by $u_\lambda^a$ and we use the antisymmetry of ${\mf A}$. We get
\begin{equation*}
\ll u_\lambda^a, g \gg = \mu \ll u_{\mu}^a, u_\lambda^a \gg + \lambda \ll u_{\lambda}^s, u_\mu^s \gg + \gamma \ll u_\lambda, (-{\mf S}) u_\mu \gg.
\end{equation*}

Since $u_{\lambda}^a \in {\mc H}^a \cap {\mc D} (\mf L)$ we have by Lemma \ref{lem:sg} that  \begin{equation*}
2 \ll u_\lambda^a , u_\lambda^a \gg \le  \| u_\lambda^a\|_1^2 = \|u_\lambda \|_1^2 - \|u_\lambda^s\|_1^2 \le C^2 \gamma^{-1}.
\end{equation*}

Remark that $u_\lambda^a$ and $u_\lambda^s$ converge weakly in ${\mathcal H}_1$ respectively to $u_0^a$ and to $u_0^s$. We first take the limit as $\lambda \to 0$ and then as $\mu \to 0$ and we obtain
\begin{equation*}
\ll u_0, g \gg = \gamma \ll u_0, (- \mf S) u_0 \gg.
\end{equation*}
On the other hand, since ${\mf S} h= g$ for a local function $h \in {\mc H}^a$ we have
\begin{eqnarray*}
\ll u_0, g \gg = - \ll u_0, h \gg_1=  - \lim_{\lambda \to 0} \ll u_\lambda, h \gg_1=\lim_{\lambda \to 0} \ll u_\lambda , g \gg \nonumber \\
=\lim_{\lambda \to 0} \left[ \lambda \ll u_\lambda, u_\lambda \gg + \ll u_\lambda, (-{\mf A}) u_\lambda \gg + \gamma \ll u_\lambda, (- {\mf S} ) u_\lambda \gg \right] \nonumber \\
=\lim_{\lambda \to 0} \left[ \lambda \ll u_\lambda, u_\lambda \gg + \gamma \ll u_\lambda, (- {\mf S} ) u_\lambda \gg \right] \nonumber \\
\geq \lim_{\lambda \to 0} \lambda \ll u_\lambda, u_\lambda \gg + \gamma \ll u_0, (-\mf S) u_0 \gg \nonumber
\end{eqnarray*}
where the last inequality follows from the weak convergence in ${ \mathcal H}_1$ of $(u_\lambda)_\lambda$ to $u_0$. It implies
\begin{equation*}
\lim_{\lambda \to 0} \lambda \ll u_\lambda, u_\lambda \gg =0
\end{equation*}
so that $u_\lambda$ converges strongly to $u_0$ in ${\mathcal H}_1$. Uniqueness of the limit follows by a standard argument.

Since $f \in {\mc H}^a$ we have ${\mf S} h =f$ for some local function $h \in{\mc H}^a$. It follows that
\begin{equation*}
\lim_{\lambda \to 0} \ll f, u_\lambda \gg = -\lim_{\lambda \to 0} \ll h ,u_{\lambda} \gg_1 = - \ll h ,u_{0} \gg_1= \ll f, u_0^a \gg.
\end{equation*}

To show that $\sigma (f,g) =\sigma (g,f)$ observe that 
\begin{equation}
\sigma (g,f) = \lim_{\lambda \to 0} \ll (\lambda- {\mf L}^*)^{-1} g, f \gg  
\end{equation}
where the adjoint ${\mf L}^*$ of ${\mf L}$ in ${\mc H}$ is given by ${\mf L}^* = -{\mf A} +\gamma {\mf S}$. We repeat the argument above with ${\mf L}$ replaced by ${\mf L}^*$ . It is easy to see that the solution $v_\lambda$ of the resolvent equation $(\lambda -{\mf L}^*) v_{\lambda}=g$ satisfies $v_{\lambda}^a = u_{\lambda}^a$, $v_{\lambda}^s = -u_{\lambda}^s$. It follows that 
$$\ll v_{\lambda} , f \gg = \ll v_{\lambda}^a , f \gg = \ll u_{\lambda}^a, f \gg = \ll u_\lambda , f \gg .$$
Taking the limit $\lambda \to 0$ we get $\sigma (g,f)= \sigma (f,g)$.
\end{proof}

\begin{prop}
\label{prop:ons} 
The Onsager relation $\kappa^{e,r}=\kappa^{r,e}=0$ holds in the following sense:
\begin{equation*}
\sigma (j_{0,1}+\tau p_0, p_0) =\sigma(p_0, j_{0,1} +\tau p_0)=0.
\end{equation*}
\end{prop}

\begin{proof}
We have
\begin{equation*}
(\lambda -{\mf L}) (p_0) =(V' (r_0) -V' (r_1)) + (2\gamma +\lambda) p_0.
\end{equation*}
Since every discrete gradient $\theta_1 u - u$ is equal to zero in ${\mc H}$ we get
\begin{eqnarray*}
 \sigma (j_{0,1}+\tau p_0, p_0) &=&\lim_{\lambda \to 0} \cfrac{1}{\lambda +2\gamma} \ll j_{0,1} +\tau p_0, p_0 \gg\\
 &=& \lim_{\lambda \to 0} \cfrac{1}{\lambda +2\gamma} \mu_{\tau,T} \left( (-p_0 V' (r_1)  + \tau p_0) p_0 \right)\\
 &=& \cfrac{-\tau T +\tau T}{2\gamma} =0
\end{eqnarray*}
because $\mu_{\tau,T} (p_0^2)=T, \; \mu_{\tau,T} (V' (r_1)) = \tau$.
\end{proof}

\section{Temperature dependance of the conductivity}
\label{sec:bounds}

The exact value of the conductivity is in general out of reach and even an estimate in terms of the parameters of the model, temperature and pressure, is difficult. The only tractable case is the harmonic chain with interacting potential $V(r)=a r^2 /2$. In this case an explicit formula is available:

\begin{equation*}
\kappa^{e} (T,\tau) = \kappa^{\rm harm}_a = \cfrac{a}{2\gamma}.
\end{equation*}

The fact that the conductivity is independent of the temperature (and the pressure)  is due to the linear interactions. For general anharmonic chain we expect a non-trivial temperature dependance. The aim of this section is to establish rigorous lower and upper bounds on $\kappa$ giving some insight on its  behavior with respect to the temperature. In the rest of this section, to simplify, we assume that pressure $\tau$ is equal to $0$ and we note $\kappa (T)$ for $\kappa^{e} (T,0)$.

The usual assumptions on $V$ are
\begin{equation*}
V(r) \sim_0 a r^2,\; a>0,  \quad V(r) \sim_{\infty} A |r|^{\alpha}, \; \alpha \ge 2, \; A>0 .
\end{equation*}


From a more general point of view it makes sense to consider situations where the potential $V$ satisfies $a=0$ or $\alpha<2$.

To get upper and lower bounds for the Green-Kubo formula a general approach is to use variational formula. We introduce ${\mc H}_{1,\lambda}$ and ${\mc H}_{-1,\lambda}$ norms defined for any smooth local function $f$ by
\begin{equation*}
\| f \|_{\pm 1,\lambda}^2 = \ll f, (\lambda -\gamma {\mf S})^{\pm 1} f \gg.
\end{equation*}

We have (see e.g. \cite{Seth}):
\begin{equation}
\label{eq:var}
\begin{split}
\ll g, (\lambda -{\mf L})^{-1} g \gg &= \sup_{f} \left\{ 2 \ll f,g \gg -\| f\|_{1,\lambda}^2 -\| {\mf A} f \|_{-1,\lambda}^2 \right\}\\
&=\inf_{f} \left\{ \| g+\mf A f \|_{-1, \lambda}^2 + \| f \|_{1,\lambda}^2 \right\},
\end{split}
\end{equation}
where the supremum (resp. infimum) is taken over a dense subset of ${\mc D} ({\mf L})$. The function $g$ we are interested in is $g=j_{0,1} $. Formula (\ref{eq:var}) is valid as soon as the operator ${\mf L}={\mf A} +\gamma {\mf S}$  is the generator of a strongly continuous Markov process in ${\mc H}$ with ${\mf A}$ (resp. ${\mf S}$) being the antisymmetric (resp. symmetric) part of ${\mf L}$. This can be easily proved if $V$ growths at most quadratically at infinity but it should be valid for a more general class of potentials. In this section we assume that potential $V$ is such that (\ref{eq:var}) is valid and that the Green-Kubo formula converges, i.e. that the conditions of Section \ref{sec:GK} are fulfilled. Moreover we assume that $V(r)>0$ if $r \ne 0$.

\subsection{Upper bounds for generic anharmonic chains}

In the sequel, ${\rm Var} (f)$ denotes the variance of the function $f$ with respect to the probability measure $\mu_{0,T}$.

\begin{prop}
\label{prop:UB}
We have the following upper bound on the conductivity
\begin{equation}
\label{eq:ub}
\kappa (T) \, \le \, \cfrac{1}{4 \gamma T} {\rm Var} (V' (r_0) ).
\end{equation}
\end{prop}

\begin{proof}
Take the function $f=- V(r_1) /2$ in the second variational formula of (\ref{eq:var}). We have 
\begin{equation*}
\begin{split}
j_{0,1} + {\mf A} f &= -p_0 V' (r_1)-\cfrac{1}{2} (p_1 -p_0) V' (r_1) \\
&=-\cfrac{1}{2} p_0 (V' (r_1) +V' (r_0)) +\cfrac{1}{2} (p_0 V' (r_0) -p_1 V' (r_1))\\
&= -\cfrac{1}{2} p_0 (V' (r_1) +V' (r_0)) 
\end{split}
\end{equation*}
because gradient terms in ${\mc H}$ are equal to $0$. 

Observe now that
\begin{equation*}
(\lambda -\gamma {\mf S})^{-1} \left[ p_0 (V' (r_1) +V' (r_0)) \right] = (\lambda + 2 \gamma)^{-1} \left[p_0 (V' (r_1) +V' (r_0)) \right]
\end{equation*} 
so that
\begin{equation*}
\| j_{0,1} + {\mf A} f \|_{-1,\lambda}^2 = \cfrac{1}{2} \cfrac{T}{\lambda +2 \gamma} {\rm Var} (V' (r_0)).
\end{equation*}
Moreover
\begin{equation*}
\| f\|_{1,\lambda}^2 = \lambda \ll f, f \gg= \lambda {\rm Var} (V (r_0)).
\end{equation*}
By taking the limit $\lambda \to 0$ we get
\begin{equation*}
\kappa (T) \, \le \cfrac{1}{4 \gamma T}  {\rm Var} (V' (r_0)).
\end{equation*}
\end{proof}

\begin{cor}
\label{cor:ub}
We have the following upper bounds on the conductivity: 
\begin{itemize}
\item {\bf{High temperature regime:}}
Assume that the smooth nonnegative potential $V$ satisfies:
\begin{itemize}
\item $V(r)=V(-r)$
\item $V(r)=Ar^\alpha +W(r), \quad r \ge 1, \quad \alpha > 1,$
\end{itemize}
with $A$ a positive constant and $W$ a smooth function such that $\| W\|_{\infty} + \| W' \|_{\infty} < + \infty$. 

Then, there exists a constant $C>0$ such that in the high temperature regime $T \to \infty$,
\begin{equation*}
\kappa (T) \le \cfrac{C}{T^{2/\alpha-1}}.
\end{equation*}

In particular, $\kappa (T)$ converges to $0$ as $T$ goes to infinity in the subharmonic regime at infinity ($1<\alpha <2$). 

\item {\bf{Low temperature regime:}}
Assume that the smooth nonnegative potential $V$ satisfies:
\begin{itemize}
\item $V(r)=V(-r)$
\item $V(r) \sim Ar^\delta, \quad r\to 0, \quad \delta \ge 2$
\item $V(r)=B r^{\alpha} +W(r), \quad r \ge 1, \quad \alpha >0,$
\end{itemize}
with $A,B$ positive constants and $W$ a smooth function such that $\| W\|_{\infty} + \| W' \|_{\infty} < + \infty$.

Then, there exists a constant $C>0$ such that in the low temperature regime $T \to 0$, 
\begin{equation}
\label{eq:ubk0}
\kappa (T) \le C T^{1-2/\delta}.
\end{equation}
\end{itemize}

In particular, $\kappa (T)$ converges to $0$ as $T$ goes to $0$ in the superharmonic regime at origin ($\delta>2$).

\end{cor}


\begin{proof}
Consider first the high temperature regime. It is easy to show that
\begin{equation*}
Z(T)= \int_{-\infty}^{\infty} dr e^{-V(r)/T} \sim 2 \left( \int_0^{\infty} e^{-A s^{\alpha}} ds \right) T^{1/\alpha}.
\end{equation*} 
To estimate ${\rm Var} (V'(r_0))$ we write
\begin{equation*}
\begin{split}
&{\rm Var} (V'(r_0))= \cfrac{2}{Z(T)} \left\{ \int_{0}^{1} (V' (r))^2 e^{-V(r)/T} dr +\int_{1}^{\infty} (A\alpha r^{\alpha -1} +W' (r))^2 e^{-V(r)/T}dr \right\}\\
&=\cfrac{2}{Z(T)} \left\{ A^2 \alpha^2 \int_{1}^{\infty} r^{2\alpha -2} e^{-V(r)/T} dr + 2A\alpha  \int_{1}^{\infty} r^{\alpha -1} W'(r) e^{-V(r)/T} dr \right.\\
&+ \left. \int_{0}^{1} (V' (r))^2 e^{-V(r)/T} dr +  \int_{1}^{\infty} (W' (r))^2 e^{-V(r)/T}dr\right\}. 
\end{split}
\end{equation*}
With the change of variables $s=r/T^{1/\alpha}$ and the fact
\begin{equation*}
e^{-V(r)/T} = e^{-A r^{\alpha} /T} (1+ {\mc O} (T^{-1})),\quad {\text{uniformly for }}  r \ge 1,
\end{equation*}
one gets 
\begin{equation*}
\begin{split}
 \int_{1}^{\infty} r^{2\alpha -2} e^{-V(r)/T} dr &\sim T^{2-1/\alpha} \int_{T^{-1/\alpha}}^{\infty} s^{2\alpha -2} e^{-A s^\alpha} ds \\
& \sim
 T^{2-1/\alpha}\, \int_{0}^{\infty} s^{2\alpha -2} e^{-A s^\alpha} ds 
\end{split}
\end{equation*}
and
\begin{equation*}
\begin{split}
& \int_{1}^{\infty} W'(r) r^{\alpha -1} e^{-V(r)/T} dr = {\mc O} (T).
\end{split}
\end{equation*}
Moreover we have
\begin{equation*}
\begin{split}
&\int_1^{\infty} (W'(r))^2 e^{-V(r)/T} dr = {\mc O} (T^{1/\alpha})
\end{split}
\end{equation*}
and the remaining term $\int_0^1 (V'(r))^2 e^{-V(r)/T} dr$ is of order ${\mc O} (1)$. The result follows. In fact we have seen that
\begin{equation}
\label{eq:Cinfty}
{\rm Var} (V'(r_0)) \sim C_{\infty} T^{2-2/\alpha}, \quad C_{\infty}= A^2 \alpha^2 \cfrac{\int_0^{\infty} s^{2\alpha -2} e^{-A s^{\alpha}} ds }{\int_0^{\infty} e^{-A s^{\alpha}} ds }.
\end{equation}

Consider now the superharmonic regime at origin and $T$ small. It is easy to show that
\begin{equation*}
Z(T)= \int_{-\infty}^{\infty} dr e^{-V(r)/T} \sim 2 \left( \int_0^{\infty} e^{-A s^{\delta}} ds \right) T^{1/\delta},
\end{equation*} 
because, taking into account the asymptotic behavior of $V$ at infinity, the part to the integral corresponding to $r \ge 1$ is exponentially small in $1/T$. Observe now that 
\begin{equation*}
\int_1^{\infty} (V' (r))^2 e^{-V(r)/T} dr ={\mc O} \left( T^{1/\alpha} \int_{T^{-1/\alpha}}^{\infty} (V'(sT^{1/\alpha}))^2 e^{-B s^{\alpha}}ds \right) ={\mc O} (e^{-B/(2T)}),
\end{equation*} 
and
\begin{equation*}
\begin{split}
&\int_0^{1} (V' (r))^2 e^{-V(r)/T} dr =T^{1/\delta} \int_0^{T^{-1/\delta}} (V'(s T^{1/\delta}))^2 e^{-V(sT^{1/\delta})} ds\\
& \sim \left( A^2 \delta^2  \int_0^{\infty} ds e^{-s^{\delta}} s^{2(\delta-1)} ds \right) \; T^{2-1/\delta}.
\end{split}
\end{equation*} 
One gets (\ref{eq:ubk0}). On the way we have seen that, as $T \to 0$,
\begin{equation}
\label{eq:C0}
{\rm Var} (V'(r_0)) \sim C_{0} T^{2-2/\delta}, \quad C_{0}= A^2 \delta^2 \cfrac{\int_0^{\infty} s^{2\delta -2} e^{-A s^{\delta}} ds }{\int_0^{\infty} e^{-A s^{\delta}} ds }\; .
\end{equation}
\end{proof}

\subsection{Lower bounds for generic anharmonic chains}

We are not able to obtain pertinent lower bounds for the conductivity of the process generated by ${\mf L}= {\mf A} + \gamma {\mf S}$ except for the exponential interaction (see below). Nevertheless, if we perturb the chain by the smoother second energy conserving noise ${\mf S}'$ considered in \cite{BO},\cite{B}, interesting lower bounds are available. Thus, we consider the infinite system with generator ${\mf L}'$ given by 
\begin{equation*}
{\mf L}' = {\mf A} +\gamma ({\mf S} + {\mf S}')
\end{equation*}   
where ${\mf S}'$ is defined by
\begin{equation}
\label{eq:gen1S}
{\mf S}'=\cfrac{1}{2}\sum_{x \in \ZZ} (p_{x+1} \partial_{p_x} -p_{x} \partial_{p_{x+1}})^2.
\end{equation}
The noise ${\mf S} +{\mf S}'$ is energy conserving and satisfies (\ref{eq:sg1}). It is not difficult to adapt the proof given in Section \ref{sec:GK} to show that
\begin{equation*}
\sigma' (j_{0,1}, j_{0,1}) = \lim_{\lambda \to 0} \ll j_{0,1} , (\lambda -{\mf L}')^{-1} j_{0,1} \gg
\end{equation*}
exists. The disadvantage of ${\mf S}'$, contrary to ${\mf S}$, is that it does not satisfy (\ref{eq:sg1}). A second difference between the noise ${\mf S}$ and the noise ${\mf S}'$ is that the latter gives a positive trivial contribution $\gamma$ to the conductivity, so that the conductivity $\kappa' (T)$ corresponding to ${\mf L}'$ is
\begin{equation*}
\kappa' (T)= \gamma + T^{-2} \sigma' (j_{0,1}, j_{0,1}).
\end{equation*}

The inequality (\ref{eq:ub}) is now replaced by
\begin{equation}
\label{eq:ub2}
\kappa' (T) -\gamma \le \, \cfrac{1}{6 \gamma T} {\rm Var} (V' (r_0) )
\end{equation}
because 
\begin{equation}
\label{eq:SSD}
({\mf S} + {\mf S}' ) p_x = -3 p_x, \quad (\mf S +{\mf S}') (p_x^2) = \Delta p_x^2.
\end{equation}

In the harmonic case $V(r)=a r^2$, we get
\begin{equation}
\label{eq:harm2}
\kappa_a^{\prime, harm} (T) =\gamma +\cfrac{a}{3 \gamma}.
\end{equation}

With this perturbation we can prove the following proposition
\begin{prop}
Consider the dynamics generated by ${\mf L}' ={\mf A} +\gamma ({\mf S} +{\mf S}')$. Then, we have
\label{prop:LB}
\begin{equation*}
\sigma' (j_{0,1}, j_{0,1}) \ge \cfrac{ T^2 \left[ {\rm Var} (V' (r_0)) \right]^2}{6\gamma T {\rm Var} (V' (r_0)) +2T^2 \gamma^{-1} {\rm Var} (V'' (r_0))}.
\end{equation*}
\end{prop}

\begin{proof}
We choose the test function 
$$v = -a  p_0 (V' (r_0) + V' (r_{1}))$$
with $a>0$ we will specify later. We have
\begin{equation}
{\mf A}  v =  a p_0^2 \left( V''(r_{1})-V'' (r_0) \right)
\end{equation}  
in ${\mc H}$ (because in ${\mc H}$ all gradient terms are equal to $0$).

Let $G_\lambda $ be the solution of the resolvent equation corresponding to the discrete Laplacian $\Delta$ on $\ZZ$:
$$ (\lambda -\gamma \Delta) G_\lambda  =\delta_0 .$$
By (\ref{eq:SSD}) we have
\begin{equation}
(\lambda -\gamma ({\mf S}+{\mf S}') )^{-1} ({\mf A} v) = a \sum_{z} G(z) p_z^2 \left( V'' (r_{1})- V'' (r_0)\right). 
\end{equation}
It follows that
\begin{eqnarray*}
\| {\mf A} v  \|_{-1,\lambda}^2 &=& a^2 \sum_{z,y} G_\lambda (z) \langle p_z^2 (V'' (r_1) -V'' (r_0))\, p_y^2 (V'' (r_{y+1}) -V'' (r_y)) \rangle\\
&=&a^2 T^2 \sum_{z \ne y} G_{\lambda} (z) \langle (V'' (r_1) -V'' (r_0))\, (V'' (r_{y+1}) -V'' (r_y)) \rangle\\
&+&3a^2 T^2 \sum_{z} G_{\lambda} (z) \langle (V'' (r_1) -V'' (r_0))\, (V'' (r_{z+1}) -V'' (r_z)) \rangle\\
&=& 2a^2 T^2 \sum_{z} G_{\lambda} (z) \langle (V'' (r_1) -V'' (r_0))\, (V'' (r_{z+1}) -V'' (r_z)) \rangle\\
&=& -2 a^2 T^2 (\Delta G_\lambda) (0) {\rm Var} (V'' (r_0)).
\end{eqnarray*}

Using again (\ref{eq:SSD}) we have (recall that the pressure is fixed to $0$ so that $V' (r_0)$ is centered)
\begin{equation*}
\|v \|_{1,\lambda}^2 = 6 a^2 \gamma T {\rm Var} (V' (r_0)), \quad \ll j_{0,1}, v \gg = aT {\rm Var} (V' (r_0)).
\end{equation*}

Hence we obtain
\begin{equation*}
\sigma' (j_{0,1}, j_{0,1}) \ge 2aT {\rm Var}  (V' (r_0)) -a^2 \left\{ 6\gamma T {\rm Var} (V' (r_0)) +2T^2 \gamma^{-1}{\rm Var} (V'' (r_0))\right\}
\end{equation*}
because 
$$\lim_{\lambda \to 0} (\Delta G_{\lambda} )(0) = -\lim_{\lambda \to 0} \int_0^1 \cfrac{4 \sin^{2} (\pi k)}{ \lambda + 4 \gamma \sin^2 (\pi k)} dk= -\gamma^{-1}.$$  
Optimizing over $a>0$ we get
\begin{equation*}
\sigma' (j_{0,1}, j_{0,1}) \ge \cfrac{ T^2 \left[ {\rm Var} (V' (r_0)) \right]^2}{6\gamma T {\rm Var}  (V' (r_0)) +2T^2 \gamma^{-1} {\rm Var} (V'' (r_0))}
\end{equation*}
which implies the result.
\end{proof}

Observe now that by an integration by parts we have
\begin{equation*}
\langle V '' (r_0) \rangle = T^{-1} {\rm Var} (V' (r_0)). 
\end{equation*}

It follows that
\begin{equation}
\label{eq:lb2}
\kappa' (T) - \gamma \ge    \left\{6\gamma \cfrac{T}{{\rm Var}  (V' (r_0))} -2/\gamma +\cfrac{2T^2}{\gamma}\cfrac{ \langle (V'' (r_0))^2 \rangle}{[{\rm Var}(V' (r_0)]^2} \right\}^{-1}.
\end{equation}

\begin{cor}
\label{cor:lb}
Consider the dynamics generated by ${\mf L}' ={\mf A}+\gamma ({\mf S}+{\mf S}')$. We have the following asymptotics for the conductivity:
\begin{itemize}
\item{\bf{High temperature regime:}}
If the potential $V$ satisfies the assumptions of corollary \ref{cor:ub} with $1<\alpha < 2$ then 
\begin{equation*}
\kappa' (T) \sim \gamma+ \cfrac{C_{\infty}}{6\gamma}\;{T^{(1-2/\alpha)}}, \quad T \to \infty, 
\end{equation*}
with $C_{\infty}$ defined in (\ref{eq:Cinfty}).

If the potential $V$ satisfies the assumptions of corollary \ref{cor:ub} with $\alpha > 2$ then
\begin{equation*}
CT^{1-2/\alpha} \ge \kappa' (T) \ge C^{-1}
\end{equation*}
with a constant positive constant $C:=C(\alpha,A)$ independent of $T$. In particular in the superharmonic regime ($\alpha >2$) the conductivity does not vanish as $T \to \infty$.

\item {\bf{Low temperature regime:}} 
Assume that the potential $V$ satisfies the assumptions of corollary \ref{cor:ub} with $2< \delta$ (superharmonic regime at origin). Then 
\begin{equation*}
\kappa' (T) \sim \gamma+ \cfrac{C_{0}}{6\gamma}\;{T^{(1-2/\delta)}}, \quad T \to 0, 
\end{equation*}
with $C_{0}$ defined in (\ref{eq:C0}).
\end{itemize}
\end{cor}

\begin{proof}
Let us start with the subharmonic regime $\alpha <2$ at infinity and let $T \to \infty$. We claim that if $\alpha>1$ then
\begin{equation}
\label{eq:comp}
T^2 \cfrac{\langle (V'' (r_0))^2 \rangle}{ {\rm{Var}}^2 (V' (r_0))} =o \left( \cfrac{T}{ {\rm {Var}} (V' (r_0)) }\right).
\end{equation}

To prove (\ref{eq:comp}) we observe that by (\ref{eq:Cinfty}) the term
$$ \cfrac{T}{ {\rm {Var}} (V' (r_0)) }$$
is of order $T^{(2/\alpha -1)}$ and that
\begin{equation*}
Z(T)^{-1} \int_0^{1} (V'' (r))^2 e^{-V(r)/T} dr ={\mc O} (1).
\end{equation*}
Hence we are left to estimate 
\begin{eqnarray*}
\int_1^{\infty} (V'' (r))^2 e^{-V(r)/T} dr &=& A^2 [\alpha(\alpha -1)]^2 \left[ \int_1^{\infty} dr r^{2\alpha -4} e^{-Ar^{\alpha}/T}\right] \left(1+ {\mc O} (T^{-1})\right)\\
&+& 4 A \alpha (\alpha -1)  \left[ \int_1^{\infty} dr r^{\alpha -2} e^{-Ar^{\alpha}/T} \right] \left(1+ {\mc O} (T^{-1})\right)\\
&+& \int_1^{\infty}  dr (W'' (r))^{2} e^{- V(r)/T} \;.
\end{eqnarray*}
The last term is trivially of order ${\mc O} (Z(T))= {\mc O} (T^{1/\alpha})$. In the two other integrals we perform the change of variables $r=T^{1/\alpha} s$. We have
\begin{equation*}
\begin{split}
&\int_1^{\infty} (V'' (r))^2 e^{-V(r)/T} dr = A^2 [\alpha(\alpha -1)]^2 T^{(2-3/\alpha)}\int_{T^{-1/\alpha}}^{\infty} s^{2\alpha -4} e^{-As^{\alpha}} \left(1+ {\mc O} (T^{-1})\right)\\
&+ 4 A \alpha (\alpha -1) T^{(1-1/\alpha)} \int_{T^{-1/\alpha}}^{\infty} s^{\alpha -2} W' (T^{1/\alpha} s) e^{-As^{\alpha}} \left(1+ {\mc O} (T^{-1})\right) + {\mc O} (T^{1/\alpha}).
\end{split}
\end{equation*}

Observe that
\begin{equation*}
\int_{T^{-1/\alpha}}^{\infty} s^{2\alpha -4} e^{-As^{\alpha}} =
\begin{cases}
{\mc O} (1), \quad \alpha >3/2,\\
{\mc O} (\log T), \quad \alpha =3/2, \\
{\mc O} (T^{-2 +3/\alpha}), \quad \alpha < 3/2,
\end{cases}
\end{equation*}
and
\begin{equation*}
\int_{T^{-1/\alpha}}^{\infty} s^{\alpha -2} e^{-As^{\alpha}} =
\begin{cases}
{\mc O} (1), \quad \alpha >1,\\
{\mc O} (\log T), \quad \alpha =1.
\end{cases}
\end{equation*}
It follows that if $1<\alpha \le 2$ we have
\begin{equation}
\int_1^{\infty} (V'' (r))^2 e^{-V(r)/T} dr ={\mc O} (T^{1/\alpha})
\end{equation}
and the claim follows.  

In the superharmonic regime $\alpha>2$ the computations are similar and the proof relies on the facts 
\begin{equation*}
\cfrac{T}{ {\rm {Var}} (V' (r_0)) } \sim C_{\infty}^{-1} T^{2/\alpha -1}, \quad T^2 \cfrac{\langle (V'' (r_0))^2 \rangle}{ {\rm{Var}}^2 (V' (r_0))} -1 \ge C
\end{equation*}
where $C$ is a positive constant independent of $T$. If $\delta > 2$ then similarly,
\begin{equation}
\label{eq:comp2}
T^2 \cfrac{\langle (V'' (r_0))^2 \rangle}{ {\rm{Var}}^2 (V' (r_0))} =o \left( \cfrac{T}{ {\rm {Var}} (V' (r_0)) }\right)
\end{equation}
as $T \to 0$ and one concludes by (\ref{eq:C0}).

%


\end{proof}

\begin{remark} In the superharmonic regime $\alpha>2$ at infinity the upper bound and lower bounds for the high temperature regime do not coincide. The $\alpha=\infty$ case formally corresponds to the Toda lattice studied in the next subsection. For the latter the upper bound is of order $1$. Hence, we conjecture that the upper bound obtained here is not sharp.
\end{remark}

If the potential is a bounded perturbation of the harmonic case then we get

\begin{cor}
Assume that the symmetric smooth potential $V$ is such that $V(r) = a r^2 +W(r)$, $a>0$, with $W, W'$ bounded, such that $W'(0)=W''(0)=0$ and $W'' (r) \to 0$ as $r \to \infty$. Then 
\begin{equation}
\kappa' (T) \sim \gamma+ {\kappa}^{\prime, harm}_a 
\end{equation}
as $T\to 0$ or $T \to \infty$, with ${\kappa}^{\prime, harm}_a$ defined by (\ref{eq:harm2}).
\end{cor}

\begin{proof}
Let us start with the high temperature regime $T \to \infty$. Recall that by (\ref{eq:Cinfty}) we have ${\rm Var} (V'(r_0)) \sim 2a T$. Moreover we have 
\begin{eqnarray*}
\langle (V'' (r_0))^2 \rangle = 2 Z(T)^{-1} \left( \int_{0}^{\infty} (2a +W'' (r))^2 e^{-ar^2 /T} dr \right)\left(1 +{\mc O} (T^{-1}) \right) 
\end{eqnarray*}
where $Z(T)$ is given by
\begin{equation*}
Z(T)=\left( 2 \int_{0}^{\infty} e^{-a r^2 /T} dr \right)\left(1 +{\mc O} (T^{-1}) \right) = {\sqrt {\cfrac{Ta}{2}} }+ {\mc O} (T^{-1/2}).
\end{equation*}
Since
\begin{equation*}
T^{-1/2} \int_{0}^{\infty} ( W'' (r))^{j} e^{-ar^2 /T} dr = \int_{0}^{\infty}  (W'' (s{\sqrt T}))^j  e^{-as^2} ds
\end{equation*}
converges to $0$ as $T \to \infty$ for $j=1,2$, we get 
\begin{equation*}
\lim_{T \to \infty} \langle (V'' (r_0))^2 \rangle = 4a^2.
\end{equation*}
Then the result follows by (\ref{eq:lb2}). We shall prove that the upper bound (\ref{eq:ub2}) and the lower bound (\ref{eq:lb2}) converges to ${\kappa}^{\prime, harm}_a$ as $T$ goes to $0$. We have
\begin{eqnarray*}
{\rm Var }(V' (r_0))&=& \cfrac{\int (2ar +W' (r))^2 e^{-(ar^2 +W(r))/T} dr}{\int e^{-(ar^2 +W(r))/T} dr}\\
&=&T \cfrac{\int (2au + T^{-1/2} W' (T^{1/2} u))^2 e^{-au^2} e^{- W(T^{1/2} u)/T} du}{\int  e^{-au^2} e^{- W(T^{1/2} u)/T} du}\\
& \sim & 4 a^2 T \cfrac{\int u^2 e^{-au^2} du}{\int  e^{-au^2} du}=2a T
\end{eqnarray*}
as $T \to 0$. Similarly we have
\begin{eqnarray*}
\langle (V'' (r_0))^2\rangle&=& \cfrac{\int (2a +W'' (r))^2 e^{-(ar^2 +W(r))/T} dr}{\int e^{-(ar^2 +W(r))/T} dr}\\
&=&\cfrac{\int (2a + W' (T^{1/2} u))^2 e^{-au^2} e^{- W(T^{1/2} u)/T} du}{\int  e^{-au^2} e^{- W(T^{1/2} u)/T} du}\\
& \sim & 4a^2
\end{eqnarray*}
as $T \to 0$. Thus, we obtain
\begin{equation}
 \lim_{T \to 0} \left\{6\gamma \cfrac{T}{{\rm Var}  (V' (r_0))} -2/\gamma +\cfrac{2T^2}{\gamma}\cfrac{ \langle (V'' (r_0))^2 \rangle}{[{\rm Var}(V' (r_0)]^2} \right\}^{-1} =\cfrac{a}{3\gamma}
\end{equation}
and
\begin{equation}
 \lim_{T \to 0} \cfrac{1}{6 \gamma T} {\rm Var} (V' (r_0) )= \cfrac{a}{3\gamma}.
\end{equation}
\end{proof}

\subsection{The Toda lattice}

The Toda lattice is the deterministic chain with generator ${\mf A}$ and asymmetric potential 
$$V(r) ={a}(e^{-r} -1) +ar.$$
The interest in this model lies in its complete integrability and its high number of conserved quantities (\cite{T}). We denote these conserved quantities by $\sum_{x \in \ZZ} \theta_x I_k$, $k \ge 1$. Let us just mention the first three ones: 
\begin{equation}
\begin{split}
&I_1 = p_0,\\
&I_2 = \left( \cfrac{p_0^2}{2} + V(r_0) \right),\\
&I_3=\left( \cfrac{p_0^3}{3} -p_0 \right) +{a}(p_0 + p_1) (e^{- r_1} -1).
\end{split}
\end{equation}

Remark that the first one corresponds to momentum conservation and the second one to energy. For an integrable model, ideal conducting behavior is expected with current correlations decaying to a finite value at long times. To estimate this limiting value, or at least obtain a lower bound (\cite{Z}), Mazur inequality (\cite{M}) is useful. It relies the long time asymptotic of dynamic correlations functions to the presence of conservation laws. Let us show how to recover a Tauberian counterpart of Mazur inequality as a simple consequence of the variational formula (\ref{eq:var}). 

In the deterministic case, the variational formula (\ref{eq:var}) is (take $\gamma =0$)
\begin{equation}
\ll g, (\lambda -{\mf A})^{-1} g \gg =\sup_{f} \left\{ 2 \ll f, g \gg - \lambda \ll f, f \gg - \lambda^{-1} \ll {\mf A} f, {\mf A}f \gg\right\}.
\end{equation}
To get a lower bound a natural idea is to use for $f$ linear combinations of the conserved quantities $I_1, \ldots,I_k,\ldots$. The term $\ll {\mf A} f, {\mf A} f \gg $ is then equal to $0$ and we get
\begin{equation}
\label{eq:mazur}
\lambda^{-1} \ll g, g \gg \; \ge \; \ll g, (\lambda -{\mf A})^{-1} g \gg \; \ge \;  \lambda^{-1} \; \ll {\mc P}g , {\mc P} g \gg
\end{equation}
where ${\mc P} g$ is the orthogonal projection on the linear space ${\mc E}$ generated by the conserved quantities.

Recall that 
\begin{equation}
\ll g, (\lambda -{\mf A})^{-1} g \gg = \int_{0}^{\infty} dt e^{-\lambda t} \ll e^{t {\mf A}} g , g \gg
\end{equation} 
where $(e^{ t {\mf A}})_{t \ge 0}$ is the semigroup generated by the Liouville operator ${\mf A}$.

Thus, if $\ll {\mc P}g , {\mc P} g \gg \, >\, 0$,  it means, in a Tauberian sense, that $ \ll e^{ t {\mf A}} g, g \gg$ remains of order $1$ as $t$ goes to infinity.  It is not difficult to see that $\ll {\mc P} j_{0,1}, {\mc P} j_{0,1} \gg\,  > \, 0$ and we recover the fact that the Toda lattice is an anomalous conductor: the conductivity defined by the Green-Kubo formula diverges. 
 
In the presence of the noise, the Toda lattice becomes a normal heat conductor. 

\begin{prop}
\label{prop:lbtoda1}
Consider the Toda lattice perturbed by $\gamma ({\mf S} +{\mf S}')$, then we have  
\begin{equation*}
\begin{split}
\gamma + \left[{ \cfrac{6 \gamma}{a} +\cfrac{2T}{\gamma a}}\right]^{-1}\; \le \; \kappa' (T) \; \le \; \gamma+ \cfrac{a}{6 \gamma}. 
\end{split}
\end{equation*}   
\end{prop}

\begin{proof}
Let $Z(T) = \int_\RR e^{-V(r)/T} dr$. Observe first that
\begin{equation*}
\cfrac{1}{a Z(T)}  \int V' (r) e^{-V(r)/T} dr =0 .
\end{equation*}
Since $V'' (r) =a- V' (r)$, by integration by parts, we have
\begin{equation}
\label{eq:stat}
\begin{split}
&\int ( V' (r) )^2 e^{-V(r) /T} dr = -T \int V' (r) \cfrac{d}{dr} (e^{-V(r)/T} ) dr\\
& = T \int V'' (r) e^{-V(r)/T} dr =T \int (a- V' (r)) e^{-V(r)/T} dr \\
&=a T Z(T) 
\end{split}
\end{equation}
and
\begin{equation*}
\langle (V'' (r_0))^2 \rangle = a^{2} -2 a \langle V'(r_0) \rangle + \langle (V' (r_0))^2 \rangle = a^2 +Ta.
\end{equation*}

We conclude by the upper bound (\ref{eq:ub2}) and the lower bound (\ref{eq:lb2}).
\end{proof}

In the same spirit as the "Tauberian  Mazur inequality" (\ref{eq:mazur}), we can obtain a lower bound on the conductivity by using elements of ${\mc E}= {\rm Span} (I_k, k\ge 1)$ as test functions. We get 
\begin{equation}
\kappa' (T) \ge \gamma + \sup_{ f \in {\mc E}} \left\{ 2 \ll f, j_{0,1} \gg - \gamma \ll f, -({\mf S} +{\mf S}') f \gg \right\}.
\end{equation} 

In the next proposition we investigate the lower bound obtained by taking $f \in {\rm Span} (I_1,I_2,I_3)$. Note that this lower bound is quite different from the previous lower bound.

\begin{prop}
\label{prop:mazurlb1}
Consider the Toda lattice perturbed by the noise $\gamma ({\mf S} +{\mf S}')$. Then, we have
\begin{equation*}
\kappa' (T) -\gamma \ge \cfrac{a^2 T^2}{ \gamma \left( 6 aT^2 +\cfrac{25}{3} T^3 -8T^2 +3T\right)}.
\end{equation*}
\end{prop}

\begin{proof}
We have
\begin{equation}
I_3=  \left( \cfrac{p_0^3}{3} -p_0 \right) +{a}(p_0 + p_1) (e^{- r_1} -1).
\end{equation} 

We inject the function $\alpha I_3$ ($\alpha >0$ will be fixed later) in the variational formula (\ref{eq:var}) with the infimum. Since ${\mf A} I_3 =0$ we get
\begin{equation*}
\kappa (T) -\gamma \ge T^{-2} \, \left\{ 2 \alpha \ll j_{0,1} ,I_3 \gg +\gamma \alpha^2 \left[\ll  {\mf S} I_3, I_3 \gg + \ll {\mf S}' I_3, I_3 \gg \right]\right\}
\end{equation*}

We have
\begin{eqnarray*}
2 \ll j_{0,1}, I_3 \gg &= & {2a^2} \ll p_0 (1-e^{- r_1}), (p_0 +p_1)(1-e^{- r_1}) \gg =2a T^2.
\end{eqnarray*}

Observe now that 
\begin{equation*}
{\mf S} I_3 = -2 I_3.
\end{equation*}
Hence $\ll {\mf S} I_3 , I_3 \gg  = -2  \ll I_3, I_3 \gg $. Using (\ref{eq:stat}) we obtain
\begin{eqnarray*}
\ll I_3, I_3 \gg &=& \ll  p_0^3 / 3 -p_0, p_0^3 / 3 -p_0 \gg \\
&+& {a^2} \ll  (p_0 +p_1) (1-e^{- r_1}), (p_0 +p_1)(1-e^{- r_1}) \gg \\
&=&  \mu_{0,T} \left\{(p_0^3 /3 -p_0)^2 \right\}+ {2 T a^2} \mu_{0, T} \left[ (1-e^{-b r_0})^2 \right] \\
&=&  \left\{ \cfrac{5}{3} T^3 -2 T^2 +T\right\} + 2a T^2.
\end{eqnarray*}

We have
\begin{equation*}
{\mf S}' I_3 = p_0 +(p_{-1} + p_{1}) p_{0} - p_{0}^3 -a (p_0 +p_1) (e^{-r_1}-1).
\end{equation*}
It follows that
\begin{eqnarray*}
\ll {\mf S}' I_3, I_3 \gg &=& -a^{2} \ll  (p_0 +p_1) (1-e^{- r_1}), (p_0 +p_1)(1-e^{-b r_1}) \gg\\
&+& \ll \left( \cfrac{p_0^3}{3} -p_0\right) , (p_0 -p_0^3 +(p_{-1} +p_1) p_0 \gg \\
&=&-\left( 2T^{2}a +5T^3 -4T^2 +T \right). 
\end{eqnarray*}

Putting everything together and optimizing over $\alpha$ we get
\begin{equation*}
\kappa (T) -\gamma \ge \cfrac{a^2 T^2}{ \gamma \left( 6 aT^2 +\cfrac{25}{3} T^3 -8T^2 +3T\right)}.
\end{equation*}
\end{proof}


Recall that the lower bound on the conductivity obtained in proposition \ref{prop:LB} was for anharmonic chains perturbed by $\gamma ({\mf S} +{\mf S}')$. For the Toda lattice we have also a lower bound even if the perturbation involves only ${\mf S}$.

\begin{prop}
Consider the Toda lattice perturbed by the noise $\gamma {\mf S}$. We have
\begin{equation*}
\kappa (T) \ge \cfrac{a^2 T^2}{2\gamma \left(\left\{ \cfrac{5}{3} T^3 -2 T^2 +T\right\} + 2a T^2 \right)}.
\end{equation*}
In particular there exists $C>0$ such that
\begin{equation*}
\begin{split}
\kappa (T) \ge CT^{-1}, \quad T \to \infty , \\
\kappa (T) \ge C T,  \quad T \to 0 .
\end{split}
\end{equation*}
\end{prop}

\begin{proof}
The proof relies on the same arguments and computations of Proposition \ref{prop:mazurlb1}.

\end{proof}

\section{Linear case}
\label{sec:harm}

In this section we assume that $V(r)=r^2 /2$.  In the bulk, i.e. for $x=2, \ldots,n-2$, we have the so-called microscopic fluctuation-dissipation equation
\begin{equation*}
j_{x,x+1} = -\cfrac{1}{4\gamma} \nabla (p_{x}^2 +r_x r_{x+1}) +{\mc L} \left[ \cfrac{r_{x+1}^2}{4} +\cfrac{(p_x +p_{x+1}) r_{x+1}}{4\gamma} \right].
\end{equation*}

It follows that 
\begin{eqnarray}
\label{eq:JJJ}
J_s&=&\langle j_{0,1} \rangle_{ss} = \cfrac{1}{n-3} \sum_{x=2}^{n-2} \langle j_{x,x+1} \rangle_{ss}\\ \nonumber
&=& -\cfrac{1}{4\gamma} \cfrac{1}{n-3} \sum_{x=2}^{n-2} \left\langle\nabla \left[ p_{x}^2 + r_{x} r_{x+1}\right]\right\rangle_{ss}\\ \nonumber
&=& \cfrac{1}{4 (n-3) \gamma} \left\{ (\langle p_2^2 \rangle_{ss} +\langle r_2 r_3\rangle_{ss} )-(\langle p_{n-1}^2 \rangle_{ss} +\langle r_{n-1} r_{n}\rangle_{ss} )\right\}.
\end{eqnarray}

We first show that this term is ${\mc O} (n^{-1})$. Our starting point is (\ref{eq:ubp}). In the following $C$ denotes a constant independent of $n$ which can change from line to line.

We have
\begin{equation}
\label{eq:eq11}
{\mc L} (p_1^2 /2) = -\tau_\ell p_1 + p_1 r_2 + \gamma_\ell (T_\ell -p_1^2), \quad {\mc L} (r_2^2 /2) = p_2 r_2 -p_1 r_2. 
\end{equation}

Since $V_s = \langle p_1 \rangle_{ss}$ is ${\mc O} (n^{-1})$ (see Lemma \ref{lem:velo}), by (\ref{eq:ubp}), we get
\begin{equation}
\label{eq:eq12}
|\langle p_1 r_2 \rangle_{ss}| \le C, \quad | \langle p_2 r_2 \rangle_{ss} | \le C.
\end{equation}

We have also
\begin{equation}
\begin{split}
&{\mc L} (p_2 r_2) = (p_2 -p_1)p_2 + (r_3 -r_2)r_2 -2 \gamma p_2 r_2,\\
&{\mc L} (p_1 r_2) =-\gamma_\ell p_1 r_2 + (r_2 -\tau_\ell) r_2 + (p_2 -p_1) p_1.
\end{split}
\end{equation}
It follows that
\begin{equation}
\label{eq:eq:13}
\langle p_2^2 +r_3 r_2 \rangle_{ss} = -2 \gamma \langle p_2 r_2 \rangle_{ss} +  \gamma_\ell \langle p_1 r_2 \rangle_{ss} + \langle p_1^2 \rangle+ \tau_\ell \langle r_2 \rangle_{ss} .
\end{equation}

By (\ref{eq:eq12}), (\ref{eq:ubp}) and Lemma \ref{lem:velo} we get
\begin{equation}
\left| \langle p_2^2 +r_3 r_2 \rangle_{ss} \right | \le C.
\end{equation}

A similar estimate can be achieved for the term $\langle p_{n-1}^2 \rangle_{ss} +\langle r_{n-1} r_{n}\rangle_{ss}$ and we obtain that
\begin{equation}
\label{eq:bJ}
\left|  J_s \right| \le C/n .
\end{equation}

\begin{theo}
Fourier law holds:
\begin{equation}
\lim_{n \to \infty} n J_s = \cfrac{1}{4\gamma} \left\{ (T_\ell -T_r) +(\tau_\ell^2 -\tau_r^2)\right\}
\end{equation}
and we have
\begin{equation}\label{eq:jthermh}
\begin{split}
{\hat J}_\ell=\lim_{n \to \infty} n(\langle p_1^2 \rangle_{ss} -T_\ell) =\cfrac{1}{4\gamma \gamma_{\ell}} \left[ (T_r -T_\ell) +(\tau_{\ell} -\tau_r)^2 \right],\\
{\hat J}_r=\lim_{n \to \infty}  n(T_r -\langle p_n^2 \rangle_{ss} )= \cfrac{1}{4\gamma \gamma_r} \left[ (T_r -T_\ell) - (\tau_{\ell} -\tau_{r})^2 \right].
\end{split}
\end{equation}
It follows that the system can be used as a heater but not as a refrigerator.
\end{theo}

\begin{proof}
By (\ref{eq:bJ}) we get that $\langle j_{0,1} \rangle_{ss} \to 0$ as $n \to \infty$. Since $V_s$ vanishes as $n$ goes to infinity we get $\langle p_1^2 \rangle_{ss} \to T_\ell$. By (\ref{eq:eq11}) it implies that $\langle p_1 r_2 \rangle_{ss}$ and $\langle p_2 r_2 \rangle_{ss}$ go to $0$. By Lemma \ref{lem:velo} and (\ref{eq:eq:13}) we have
\begin{equation}
\langle p_2^2 +r_3 r_2 \rangle_{ss} \to (T_\ell +\tau_\ell^2).
\end{equation}
Similarly one can prove
\begin{equation}
\langle p_{n-1}^2 +r_n r_{n-1} \rangle_{ss} \to (T_r +\tau_r^2).
\end{equation}
We report in (\ref{eq:JJJ}) and we get Fourier law.

Assume that $T_r > T_{\ell}$.  The term ${\hat J}_\ell$ (resp. ${\hat J}_r$) is the macroscopic heat current from the left reservoir to the system (resp. from the system to the right reservoir). Whatever the values of $\tau_\ell, \tau_r$ are, ${\hat J}_{\ell} >0$ and we can not realize a refrigerator.  But if $(T_r -T_{\ell}) < (\tau_r -\tau_{\ell})^2$ then ${\hat J}_r <0$ and we realized a heater.
\end{proof}

\section{Existence and uniqueness of the non equilibrium 
stationary   state} 
\label{sec:exist}

The aim of this section is to prove Proposition \ref{prop:car22}. We recall that $\Omega_n= \RR^{n-1} \times \RR^n$ is the state space. For any positive $\theta$ we define the Lyapunov function $W_\theta$ by
\begin{equation*}
W_{\theta} (\omega) =\exp \left( \theta \sum_{x=1}^{n} {\mc E}_x \right), \quad \omega \in \Omega_n,
\end{equation*} 
and the weighted Banach space
\begin{equation*}
\BB_{\theta} = \left\{ f: \Omega_n \to \RR \text{ contiuous }, \; \| f\|_{\theta} = \sup_{\omega} \cfrac{| f(\omega)|}{W_{\theta} (\omega)} < +\infty \right\}.
\end{equation*}

Let $F_x$ be the flip operator defined by $(F_x f)(\omega)= f(\omega^x)$ for any $f:\Omega_n \to \RR$ and any $\omega \in \Omega_n$. We note $(T_t)_{t \ge 0}$ the semigroup generated by ${\mc L}$ and $({\tilde T}_t)_{t \ge 0}$ the semigroup corresponding to ${\mathcal L}_{0}$, defined by (\ref{eq:L0}).

The existence and uniqueness of the stationary state can be proved using similar arguments as in \cite{C}, \cite{RB}. Nevertheless we are not able to show smoothness results for the transition probabilities of $(T_t)_{t \ge 0}$. 

We assume that the potential $V$ satisfies (\ref{eq:hyp}). In \cite{C} is investigated the problem of existence and uniqueness of the stationary state for ${\tilde T}_t$ in the case $\tau_\ell =\tau_r =0$. It is easy to adapt the proof of \cite{C} when $\tau_\ell, \tau_r \ne 0$. In this case, without the jumps, we can apply directly the method of \cite{C} to obtain also the smoothness of the density w.r.t. the Lebesgue measure.

\begin{prop}
\label{prop:car}
Fix $\theta$ sufficiently small. The semigroup $({\tilde T}_t)_{t \ge 0}$ can be extended to a strongly Feller continuous semigroup on $\BB_{\theta}$ with a smooth density w.r.t. the Lebesgue measure. It has a unique invariant probability measure $\pi$ which has a smooth density w.r.t. the Lebesgue measure and the semigroup converges exponentially fast to $\pi$ in $\BB_{\theta}$. 
\end{prop}

A simple computation shows that
\begin{equation}
\label{eq:sss}
\begin{split}
{\mc L} W_{\theta} &= \theta W_{\theta} \left\{ \tau_r p_n -\tau_\ell p_1 + T_{\ell}  + T_r + (T_\ell \theta -1) p_1^2 + (T_r \theta -1)p_n^2 \right\} \\
&\le \theta \left( T_r + T_\ell + \cfrac{\tau_r^2}{4(1- T_r \theta)} + \cfrac{\tau_\ell^2}{4(1- T_\ell \theta)} \right)  W_{\theta}
\end{split}
\end{equation}
if $\theta < \min (T_\ell^{-1}, T_r^{-1})$. It follows that for such a choice for $\theta$ the semigroup $(T_t)_{t \ge 0}$ is well defined on $\BB_{\theta}$. 

Similarly to Lemma 7.1 in \cite{C} one can show that there exists $t_0 >0$, constants $b_n < +\infty$, $0 < \kappa_n <1$, with $\lim_{n \to \infty} \kappa_n =0$, and compact sets $K_n$ such that
\begin{equation}
\label{eq:ttt}
T_{t_0} W_{\theta} (\omega) \le \kappa_n W_{\theta} (\omega) + b_n {\bf 1}_{K_n} (\omega)
\end{equation}

This is sufficient to apply Theorem 8.9 in \cite{RB} and get the existence of a stationary state. 

Let us also mention that  (\ref{eq:ttt}) and (\ref{eq:sss}) imply
\begin{equation}
\label{eq:ddd}
\sup_{t \ge 0}\;  (T_t W_{\theta}) (\omega) \le C( W_{\theta} (\omega) +1) 
\end{equation}
for a positive constant $C$. This is because (\ref{eq:ttt}) gives for any $p \ge 1$ that
\begin{equation*}
(T_{pt_0} W_{\theta}) (\omega) \le \kappa_1^p W_{\theta} (\omega) + \cfrac{b_1}{1-\kappa_1}
\end{equation*}
and (\ref{eq:sss}) gives the existence of a constant $c>0$ such that
\begin{equation*}
(T_s W_{\theta}) (\omega) \le e^{cs} W_{\theta} (\omega), \quad 0 \le s \le t_0,
\end{equation*}
so that, writing $T_t =T_{t - p t_0} \circ T_{pt_0}$ with $pt_0 \le t < pt_0 +1$, we obtain (\ref{eq:ddd}).

We now consider the problem of uniqueness of the steady state. 

Let us denote ${\tilde p}_t (\omega,\xi) d\xi$ the probability transition corresponding to $({\tilde T}_t)_{t \ge 0}$ and ${p}_t (\omega, d\xi)$ the probability transition corresponding to $(T_t)_{t \ge 0}$.

Let $\sigma_1$ be the stopping time defined as the first time a momentum is flipped. Observe that $\sigma_1$ has an exponential law of parameter $n\gamma$. For every bounded measurable function $f:\Omega_n \to \RR$, we have
\begin{equation}
\label{eq:sgi}
\begin{split}
&(T_t f) (\omega)= {\mathbb E}_{\omega} \left[ f(\omega (t)) {\bf 1}_{\sigma_1 <t}\right] + {\mathbb E}_{\omega} \left[ f(\omega (t)) {\bf 1}_{\sigma_1  \ge t}\right]\\
&= e^{-\gamma nt} \int_{\xi \in \Omega_n} {\tilde p}_{t} (\omega, \xi) f(\xi) d\xi \\
&+ {\gamma} \int_0^t ds e^{-\gamma ns}\;  {\sum_{x=1}^{n}\, \int_{\xi \in \Omega_n} d\xi\, {\tilde p}_s (\omega, \xi) \left( \int_{\xi' \in \Omega_n } p_{t-s} (\xi^x, d \xi') f(\xi') \right)\,}.
\end{split}
\end{equation}

We can iterate the argument and we obtain the following formula for $p_t$

\begin{equation*}
\begin{split}
&p_t (\omega, d\xi) = e^{-\gamma n t} {\tilde p}_{t} (\omega, \xi) +\\
&\sum_{k=1}^{\infty}  {\gamma^k} \sum_{x_1, \ldots, x_k=1}^{n} \left[ \int_0^\infty \ldots \int_0^\infty  ds_1 \ldots ds_{k+1}  e^{-\gamma n (s_1 +\ldots +s_{k+1})} {\bf 1}_{\{ s_1 + \ldots +s_k \le t < s_1 + \ldots s_{k+1}\}} \right.\\
& \left. \int_{\xi_1, \ldots ,\xi_k  \in \Omega^k_n}  {\tilde p}_{s_1} (\omega, \xi_1) {\tilde p}_{s_2} (\xi^{x_1}_1, \xi_2) \ldots {\tilde p}_{s_k} (\xi^{x_{k-1}}_{k-1}, \xi_k) {\tilde p}_{t-(s_1 +\ldots s_k)} (\xi_k^{x_k} , \xi) d\xi_1 \ldots d\xi_k \right].
\end{split}
\end{equation*}

This shows that $p_t (\omega, d\xi)=p_t (\omega,\xi)d\xi $ is absolutely continuous with respect to the Lebesgue measure on $\Omega_n$. Therefore, $(T_t)_{t \ge 0}$ is strongly irreducible, i.e. for every $t>0$, every $\omega \in \Omega_n$ and every open subset $O$ of $\Omega_n$,
\begin{equation*}
p_t (\omega, O) >0
\end{equation*}
because ${\tilde T}_t$ is strongly irreducible.

\begin{lemma}
The semigroup $(T_t)_{t \ge 0}$ is strongly Feller, i.e. it maps bounded measurable functions to continuous bounded functions.
\end{lemma}

\begin{proof}
By Proposition \ref{prop:car},  the semigroup $({\tilde T}_t)_{t \ge 0}$ is strongly  Feller.  It implies (see e.g. Corollary 2.4 of \cite{SW}) that for every $t>0$, for all compact sets $K \subset \Omega_n$, we have
\begin{equation}
\label{eq:sf}
\lim_{\delta \to 0} \; \sup_{\substack{|\omega - \omega'| \le \delta,\\ \omega, \omega' \in K}}\;  \sup_{\| u \|_{\infty} \le 1} \left| ({\tilde T}_t u)(\omega) -({\tilde T}_t u )(\omega') \right| =0.
\end{equation}

Let $f$ be a bounded measurable function with $\| f \|_{\infty} \le 1$. We have to show that, for any fixed $t>0$, $T_t f$ is a continuous bounded function. By (\ref{eq:sgi}), we have
\begin{equation}
\begin{split}
&(T_{t} f)(\omega') -(T_t f)(\omega)=e^{-\gamma n t} \left(  ({\tilde T}_t f)(\omega')-  ({\tilde T}_t f)(\omega)  \right)\\
&+\gamma \sum_{x=1}^{n} \int_{0}^t  e^{-\gamma n (t-s)} \left\{ \left( {\tilde T}_{t-s} \circ F_x \circ T_s \circ f\right) (\omega) -\left( {\tilde T}_{t-s} \circ F_x \circ T_s \circ f \right) (\omega')\right\} ds.
\end{split}
\end{equation}
Observe that the absolute value of  the second term on the right hand side is bounded above by 
\begin{equation*}
\gamma \sum_{x=1}^n \int_0^t e^{-n\gamma (t-s)} \; \sup_{\| g \|_{\infty} \le 1} \left| \left( {\tilde T}_{t-s} g\right) (\omega) - \left( {\tilde T}_{t-s} g\right) (\omega')  \right| \; ds 
\end{equation*} 
because 
\begin{equation*}
\| F_x \circ T_s \circ f \|_{\infty} =\sup_\xi \left| {\bb E}_{\xi^x} (f (\omega_s ) \right| \le \|f \|_{\infty} \le 1.
\end{equation*}

By the bounded convergence theorem and (\ref{eq:sf}) we have 
\begin{equation*}
\lim_{\omega' \to \omega} ( \, (T_t f)(\omega') - (T_t f)(\omega) \, )=0. 
\end{equation*}
\end{proof}

These two last properties (irreducibility and strong Feller property) are sufficient to have uniqueness of the invariant measure $\mu_{ss}$. To show that the latter has a density, we observe that for any $t>0$, the condition $\mu_{ss} T_t =T_{t}$ means that for any measurable set $A$ of $\Omega_n$ we have
\begin{equation*}
\begin{split}
&\mu_{ss} (A)= \int_{\Omega_n} d\mu_{ss} (\omega) \left( \int_{\Omega_n} {\bf 1}_A (\xi) p_{t} (\omega, \xi) d\xi \right)\\
&= \int_{\Omega_n} {\bf 1}_A (\xi) \left( \int_{\Omega_n} d\mu_{ss} (\omega) p_{t} (\omega, \xi) \right) d\xi 
\end{split}
\end{equation*}
where the second line follows from Fubini theorem.

\end{document}